\documentclass[10pt,twocolumn,secnumarabic,amssymb,nobibnotes, aps,prd,superscriptaddress, longbibliography]{revtex4-2}

\usepackage[T1]{fontenc}
\usepackage{amssymb}
\usepackage{amsmath}
\usepackage{physics}
\usepackage{multirow}
\usepackage{float}

\usepackage{algorithm}
\usepackage{algpseudocode}

\usepackage{graphicx}
\usepackage{subcaption}
\usepackage{hyperref}
\hypersetup{
    colorlinks=true,
    linkcolor=blue,
    citecolor=blue,
    urlcolor=blue,
}

\begin{document}

\title{Probing the Ground State of the Antiferromagnetic Heisenberg Model on the Kagome Lattice using Geometrically Informed Variational Quantum Eigensolver}

\author{Abdellah Tounsi}
\affiliation{Constantine Quantum Technologies, \\Fr\`{e}res Mentouri University Constantine 1, Ain El Bey Road, Constantine, 25017, Algeria}
\affiliation{Laboratoire de Physique Math\'{e}matique et Subatomique, Fr\`{e}res Mentouri University Constantine 1, Ain El Bey Road, Constantine, 25017, Algeria}%
\author{Nacer Eddine Belaloui}
\affiliation{Constantine Quantum Technologies, \\Fr\`{e}res Mentouri University Constantine 1, Ain El Bey Road, Constantine, 25017, Algeria}
\affiliation{Laboratoire de Physique Math\'{e}matique et Subatomique, Fr\`{e}res Mentouri University Constantine 1, Ain El Bey Road, Constantine, 25017, Algeria}%
\author{Abdelmouheymen Rabah Khamadja}
\affiliation{Constantine Quantum Technologies, \\Fr\`{e}res Mentouri University Constantine 1, Ain El Bey Road, Constantine, 25017, Algeria}
\affiliation{Laboratoire de Physique Math\'{e}matique et Subatomique, Fr\`{e}res Mentouri University Constantine 1, Ain El Bey Road, Constantine, 25017, Algeria}%
\author{Takei Eddine Fadi Lalaoui}
\affiliation{Constantine Quantum Technologies, \\Fr\`{e}res Mentouri University Constantine 1, Ain El Bey Road, Constantine, 25017, Algeria}
\affiliation{Laboratoire de Physique Math\'{e}matique et Subatomique, Fr\`{e}res Mentouri University Constantine 1, Ain El Bey Road, Constantine, 25017, Algeria}%
\author{Mohamed Messaoud Louamri}
\affiliation{Constantine Quantum Technologies, \\Fr\`{e}res Mentouri University Constantine 1, Ain El Bey Road, Constantine, 25017, Algeria}
\affiliation{Theoretical Physics Laboratory, University of
Science and Technology Houari Boumediene, BP 32 Bab Ezzouar,
Algiers, 16111, Algeria}

\author{David E. {Bernal Neira}}
\affiliation{Davidson School of Chemical Engineering, Purdue University, 480 Stadium Road, West Lafayette, IN, 47907, USA}

\author{Mohamed Taha Rouabah}
\email{m.taha.rouabah@umc.edu.dz}
\affiliation{Constantine Quantum Technologies, \\Fr\`{e}res Mentouri University Constantine 1, Ain El Bey Road, Constantine, 25017, Algeria}
\affiliation{Laboratoire de Physique Math\'{e}matique et Subatomique, Fr\`{e}res Mentouri University Constantine 1, Ain El Bey Road, Constantine, 25017, Algeria}%
\affiliation{Department of Electrical and Computer Engineering, North Carolina State University, Raleigh, NC 27606, USA}

\begin{abstract}
    This work investigates the nature of the ground state of the antiferromagnetic Heisenberg model on fundamental kagome cells, a triangle and a star, using the variational quantum eigensolver~(VQE) algorithm on real quantum hardware. We demonstrate that the ground state preparation is achievable using a shallow hardware-efficient quantum circuit with a naturally Euclidean parameter space. Our custom ansatz is capable of accurately recovering meaningful properties of the ground state such as the spin-spin correlation terms and static structure factor without explicit error mitigation. These features are found to be resilient to noise. We exploited the Fubini-Study metric in constructing the ansatz, ensuring a singularity-free parameter space.
    With this ansatz design, the adaptive gradient descent optimizer achieves a faster convergence in terms of the number of iterations compared to simultaneous perturbation stochastic approximation~(SPSA). We further apply error mitigation techniques, including zero-noise extrapolation (ZNE) and qubit-wise readout error mitigation (REM). While ZNE does not obey the Rayleigh-Ritz variational principle, the conditions under which REM preserves it are discussed.
\end{abstract}
\maketitle
\section{Introduction}
\label{sec:introduction}
Finding the ground-state energy of a $k$-local Hamiltonian is known to be a quantum Merlin--Arthur hard (QMA-hard) problem~\cite{kempe2004complexity}, the quantum analog of a classical NP-hard problem. This is mainly due to the exponential growth of the Hilbert space in interacting quantum many-body systems, rendering the problem generally intractable for classical computers.
Despite QMA-hard problems being challenging to solve even on a perfect quantum computer, the latter remains the most suitable tool to crack this class of problems because it follows the same laws of quantum mechanics~\cite{Feynman1982, McArdle2020, Nielsen&Chuang2010}.
However, despite recent advancements in the technology~\cite{Somoroff2023,Ganjam2024}, current quantum computers are yet to reach their promised potential to efficiently simulate complex quantum systems.
Short coherence times, high levels of noise, and limited connectivity, all characteristics of \emph{noisy intermediate-scale quantum}~(NISQ) era computers~\cite{preskill2018quantum,Bharti2022}, still pose significant challenges to quantum advantage.
Furthermore, identifying and preparing the ground state is crucial for studying many-body systems, enabling the understanding of their properties in the low-temperature regime.
For this task, the variational quantum eigensolver~(VQE) was introduced~\cite{Peruzzo2014, Nacer2025}. It is a hybrid quantum-classical algorithm specifically designed to approximate the ground-state energy using parametrized quantum circuits that are classically optimized.
Thanks to its flexibility and hybrid nature, this algorithm offers several potential advantages.
First, it can be scalable with a suitable choice of ansatz, thus handling larger systems with a lower classical computational cost, effectively reducing the computational cost associated with the growth of the Hilbert space.
Second, the VQE is designed to be compatible with current NISQ devices~\cite {Cao2019, Peruzzo2014, Cerezo2021} and their low quantum volume~\cite{Bharti2022}.
This is because it relies on relatively shallow quantum circuits to prepare trial states, and the process of observable measurements is based on primitive sampling.
Nevertheless, estimating the energy with this algorithm comes at the cost of lower accuracy compared to quantum phase estimation~(QPE)~\cite{Kitaev1995}, a trade-off that is necessary in the absence of the fault-tolerant quantum computers which are required to deal with the deep circuits of QPE.
Different methods have been developed to address this accuracy gap with NISQ computers.
Multiple error mitigation~\cite{cai2023quantum} and measurement optimization schemes~\cite{Yi2024} have been proposed to improve the energy estimates during the VQE and its scalability, while alternative methods, such as quantum sampling-based algorithms~\cite{robledo2025chemistry, kanno2023quantum}, have shown better scalability and a higher resilience to noise.

This paper is interested in the spin-$\frac{1}{2}$ isotropic Heisenberg antiferromagnetic model on a kagome lattice, a geometric arrangement of spins that has attracted increasing interest among researchers~\cite{Balents2010, Semeghini2021, Bosse2022, Kattemolle2022, Sajjan2023, guo2025quantum, Tim2025, Wang2025} for its exotic magnetic properties and potential to manifest quantum spin liquids~\cite{Balents2010}, or topological phases~\cite{Balents2010, Semeghini2021} which are still promising candidates to perform quantum computation~\cite{Bauer2014,Mei2017,tounsi2024}.
Due to the frustrated nature of this system, preparing its ground state is non-trivial~\cite{Mendels2011, Balents2010}.\\
\begin{figure}
    \centering
    \includegraphics{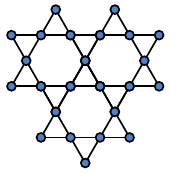}
    \caption{The geometry of the kagome lattice. }
    \label{fig:kagome-lattice}
\end{figure}
The kagome lattice is made up of corner-sharing triangles as shown in Fig.~\ref{fig:kagome-lattice}. When spins are placed on each node, the interaction between spins can be modeled quantum mechanically by the Hamiltonian of the Heisenberg interaction, $\sum_{ij} J_{ij}\mathbf{S}_i \cdot \mathbf{S}_j$, where the total spin is a conserved quantity~\cite{Teixeira2021}. Although the ferromagnetic regime ($J_{ij} < 0$) is well understood in the literature, the antiferromagnetic regime ($J_{ij} > 0$) is very intriguing since the triangular connections make it impossible to satisfy all constraints simultaneously.

Specifically, we are interested in an isotropic antiferromagnetic spin-$1/2$~Heisenberg model:
\begin{equation}
\label{eq:hamiltonian}
H = \sum_{\langle i, j \rangle} {X}_i{X}_j + {Y}_i {Y}_j + {Z}_i {Z}_j,
\end{equation}
where $\langle i, j\rangle$ is a pair of coupled sites $i$ and $j$. The Hamiltonian forms a commuting set of operators with $\mathbf{S}^2$ and $S_z$, such that $\mathbf{S} = \sum_i \mathbf{S}_i$, where $\mathbf{S}_i = ({S_{x}}_i,{S_{y}}_i,{S_{z}}_i)^T$, is the total spin vector operator and $S_z = \sum_i {{S}_z}_i = \frac{\hbar}{2}\sum_i {Z}_i$ is the $z$-component of the total spin vector. That is, $H = \sum_{i j} J_{ij}\mathbf{S}_i\cdot\mathbf{S}_j$, such that $J_{ij} = 4/\hbar^2$ if $i$ and $j$ are nearest neighbors to each other and $J_{ij}=0$ otherwise. Moreover, the Hamiltonian along with the total spin operator commutes with the lattice symmetry operations. Generally, the Hamiltonian commutes with $\mathbf{S}$, as shown in Appendix~\ref{app:block-diagonal-fubiny-study}. Therefore, the Hamiltonian commutes with the $S_{\pm}$ ladder operators, and it is SU(2) invariant.\\
For a Heisenberg model in a lattice of simply connected triangles, where it is possible to form a singlet state in each triangle, we may show, following the same arguments as \citet{shastry1981}, that a dimer state $\ket{\Psi^n_\text{dimer}}$ is a ground state. The latter is defined as such:
\begin{equation}
    \ket{\Psi^n_\text{dimer}} = \bigotimes_{\langle i, j \rangle \in \mathcal{D}_n} [i, j],
\end{equation}
where $[i, j] = \frac{1}{\sqrt{2}} \left( \ket{\uparrow_i \downarrow_j} - \ket{\downarrow_i \uparrow_j} \right)$ for each neighbor pair $\langle i, j \rangle$ in a dimer configuration $\mathcal{D}_n$ from the set of all possible dimer configurations $\mathcal{D}$.\\
Examples of simply connected triangles are the simple one-triangle, two connected triangles, one kagome star, and a chain of kagome stars. However, the argument fails for an arbitrary kagome lattice, such as three overlaid kagome stars, as shown in Fig.~\ref{fig:kagome-lattice}, because the configuration becomes geometrically frustrated and the condition of forming a singlet state in each triangle is violated. For that reason, the resonating valence bond (RVB) spin liquid state is proposed, which features short-range correlation and gapped excitations~\cite{Sachdev1992}.
However, numerical studies showed signs of long-range correlations with algebraic decay and gapless spectrum~\cite{Waldtmann1998}. For instance, using the Gutzwiller projected wave function, it has been shown that U(1) Dirac spin liquid has better agreement with exact diagonalization for $48$ sites~\cite{Ran2007}.\\
This system's general ground state is not fully understood yet, especially in the thermodynamic limit, i.e., an infinite lattice. Interestingly, numerical studies show very distinct results; initial numerical approaches of the density-matrix renormalization group (DMRG) showed a gapped ground state of $\mathbb{Z}_2$ spin liquid with short-range entanglement~\cite{Simeng2011}, while recent DMRG studies hint at algebraic states such as the $U(1)$ Dirac spin liquid~\cite{He2017} or chiral spin liquid~\cite{Sun2024}.
Moreover, \citet{Liao2017} used the projected entangled simplex states (PESS) to show that the ground state exhibits gapless quantum spin liquid in the infinite lattice limit.
Exact diagonalization methods showed that the ground state differs from the theoretical candidates, at least in the small finite sizes of 36 and 48 spin sites~\cite{Lauchli2019}. Recently, a physically oriented neural network ansatz has been employed for lattices of 48 sites and showed signs of a spinon pair density wave that does not break time reversal symmetry or any of the lattice symmetries~\cite{Tania2025}.

Quantum experiments on the kagome lattice are still very limited.
Theoretical proposals and classical simulations have been carried out. Some studies are devoted to optimizing the implementation of the Hamiltonian variational (HV) ansatz~\cite{Bosse2022, Kattemolle2022}, while some studies like \citet{guo2025quantum} utilized the HV ansatz and showed phase transitions when considering different ratios of second-nearest neighbor interactions to first-nearest neighbor ones.
Although the HV ansatz is physically inspired, it is still too deep for NISQ devices.

To motivate our approach, we compare it to recent state-of-the-art efforts that explore similar directions from different starting points.
\citet{Sajjan2023} follow a hybrid classical-quantum pipeline: they begin with DMRG computations on a reduced Hamiltonian to extract correlation data, then design an ansatz to replicate those features, and finally simulate and implement the VQE circuit, complemented by error mitigation techniques like the twirled readout error extinction (T-REx) and the zero-noise extrapolation (ZNE).
However, their implementation ultimately reduces the whole 12-qubit problem to an effective two-qubit ansatz.
Although this reduction is well motivated, it does not offer a useful benchmark for our setting, where our objective is to evaluate the VQE procedure under more problem-agnostic and less informed conditions.
Conversely, we design a generic ansatz based on the hardware connections and native gates, then adapt it to give at least one ground state and analyze it with the Fubini-Study metric to avoid zero metric or undefined geometry in the energy landscape.
Despite the presence of noise, we are able to characterize the resulting quantum state and extract structural information.
Notably, our optimization starts from product states of high energy and still converges, highlighting the robustness of the variational landscape and the optimization strategy.
\citet{Tim2025} use a symmetry-based qubit reduction strategy, identifying exact and approximate $\mathbb{Z}_2$ symmetries to taper the Hamiltonian down to five qubits.
After validating the spectrum via DMRG, they apply VQE with extensive error mitigation, including the readout-error mitigation (REM), multi-fold ZNE, and symmetry verification of measurement outcomes.
While their method excels at preserving energy estimates under strong qubit compression, its scope is restricted to the energy alone.
Our aim, by contrast, focuses on the structure and correlations of the quantum state itself, which is essential for understanding ordering, entanglement, and potential quantum phases beyond the spectral analysis.
\citet{Semeghini2021} follow a different experimental direction by preparing the resonating valence bond (RVB) states using Rydberg atom arrays. They encode each edge of the kagome lattice as an atom and use adiabatic evolution with the blockade mechanism's constraints to prepare dimer coverings, enabling the study of topological properties and quasiparticle excitations.
Despite the scalability of their platform and its suitability to probe RVB physics, it is confined to a specific candidate state and does not address the general solution of the Heisenberg model.

In this work, we focus on the ability to characterize the ground state in current quantum devices. We will show that it is possible to probe the spin-spin correlations with sufficient accuracy despite noise.
 To prepare the ground state, we design a custom shallow hardware-efficient ansatz targeting two systems: a single triangle and six corner-sharing triangles making one kagome star—geometries whose ground states correspond to simple dimer coverings and are thus product states.
We employ the quantum natural gradient descent (QNGD), an optimization technique designed to linearly transform gradient updates using the Fubini-Study metric tensor~\cite{Stokes2020}, thereby accounting for the local geometry of the parameter space. Recently, it has been shown that the QNGD method is able to outperform the vanilla gradient descent method on Rydberg atoms and superconducting circuits platforms using realistic simulations~\cite{Federico2025}. To further enhance convergence, \citet{Atif2023} developed the adaptive-QNGD (AQNGD) that combines QNGD with a backtracking line search to dynamically adapt the step-size. Instead of explicitly implementing the QNGD optimization, we construct the ansatz such that the Fubini-Study metric is implicitly diagonal and constant. Because of this, the AQNGD is reduced to the adaptive gradient descent (AGD) and any optimizers used with this ansatz would be equivalent to its quantum natural implementation.

In addition to this setting, error mitigation with low postprocessing cost is applied at the end of the VQE to improve the final ground energy estimate.
We implement the readout-error mitigation (REM) to reverse the SPAM errors using a partitioned response matrix approach~\cite{Geller2021}. Rather than reconstructing the full response matrix, we apply the partitioned inverse response matrices directly to the measured bitstrings. For sparse VQE measurements where $M \ll 2^n$ unique bitstrings are observed (with $n$ qubits), this approach scales as $O(M \times 2^n)$ in the worst case, with the possibility of truncation-based enhancement, making an improvement over the work of \citet{Geller2021} that requires $O(2^n \times 2^n)$ full matrix multiplication for post-processing. That is detailed in Appendix~\ref{sec:rem}.\\
Beyond the energy estimation, we are able to characterize the dimer state by measuring the spin-spin correlations and the static structure factor, which exhibits resilience to noise. We provide the static structure factor map with and without REM, and show a plausible agreement with the ground-state static structure factor.\\

In the following Sec.~\ref{sec:methods}, we outline the quantum natural gradient based optimization method and its implementation in Sec.~\ref{sec:aqngd}, the design of the hardware-efficient Euclidean ansatz circuit in Sec.~\ref{sec:ansatz}. The grouping of Pauli terms for efficient energy measurement along with measurement variance and standard error expectation is detailed in Appendix~\ref{sec:qwc-pauli}. Readout error mitigation and the zero-noise extrapolation methods are detailed in Appendix~\ref{sec:rem} and Appendix~\ref{sec:zne}, respectively. The results of experiments on one triangle and a single kagome star are detailed in Sec.~\ref{sec:results}, while the main discussion of findings can be found in Sec.~\ref{sec:discussion}, and the conclusion in Sec.~\ref{sec:conclusion}.
\section{Methods} \label{sec:methods}
In general, the parameters of the quantum circuit in variational quantum algorithms are optimized similarly to other classical optimization problems in machine learning models. Nevertheless, the quantum nature of the circuit imposes some genuine challenges. To begin with, the quantum measurement outcome is intrinsically probabilistic even in the presence of a noise-free ideal quantum computer. Moreover, current quantum devices suffer from additional noise sources coming from coherent and incoherent errors, state preparation errors, and measurement errors. In addition, the ansatz requires a significant number of parameters to cover the Hilbert space region that includes the ground state.\\
The cost to estimate the gradient in gradient-based optimizers is proportional to the number of parameters $d$; it is still possible, though, to optimize the number of shots for gradient estimation using the parameter-shift rule~\cite{Mitarai2018,Schuld2019}.
Finally, the cost function landscape is complex. Besides local minima, variational quantum algorithms can also face the problem of barren plateaus, when the gradient magnitude vanishes exponentially with the system's size~\cite{larocca2025barren}.\\
Given these properties, the VQE algorithm must be designed with careful considerations of the hardware capabilities and the problem's characteristics to design a well-suited ansatz--optimizer pair.
\subsection{Adaptive Quantum Natural Gradient Descent (AQNGD)}
\label{sec:aqngd}
Searching for the correct parameters that represent the ground state is the central challenge for the classical part of VQE. For this task, various optimizers have been used, such as the normal gradient descent and stochastic perturbative ones. However, they do not take into account the geometry of the space of parametrized quantum states, called the K\"ahler manifold or the projective quantum space, considering it to be represented by a Euclidean space that is free of singularities. In quantum applications, the space of parametrized quantum states is rarely Euclidean~\cite{Kolodrubetz2017,Bukov2019}; it usually has a curvature featured by the Fubini-Study metric and hosts singularities and covariant parameters, which make the search for the optimal parameters very challenging~\cite{yamamoto2019natural}. Following the natural gradient approach from classical machine learning~\cite{amari1998natural,Stokes2020}, the quantum natural gradient descent (QNGD) optimizer was proposed as its quantum analog. This optimizer uses the steepest gradient descent in the parameter space with an induced Fubini-Study metric, successfully avoiding singularities and considering parameters' covariance. The transformation from the Euclidean space to the Riemannian parameter space is done by taking the inverse of the Fubini-Study metric tensor, which is the real part of the quantum Fisher information matrix. The Fubini-Study metric can also be used, as we will show below, to decide on an ansatz design with a naturally Euclidean parameter space that is more easily trainable at a lower cost than more complex ones. For more details on the QNGD, see Appendix~\ref{app:block-diagonal-fubiny-study} and the works of \citet{yamamoto2019natural} and \citet{Stokes2020}.\\
Although QNGD trains parameters faster and in the correct convergence direction, it still faces the problem of choosing the ideal learning rate, the algorithm can get stuck around the ground state and never reach the minimum if the geometry becomes tighter than the step size, which is common in entangled systems with a large learning rate~\cite{yamamoto2019natural}. On the other hand, choosing an extremely small step size will significantly slow the algorithm since it will be making really small steps when the geometry does not necessitate it.
Choosing a step size that monotonically decreases is not a reliable solution either as it is not geometrically informed and will be heavily problem dependent.\\
A way around this was presented in~\citet{Atif2023}, who used QNGD with a variable, geometrically informed, learning rate following the Armijo rule, which is an efficient backtracking rule with proven convergence. This allows us to adaptively choose a large step size when needed to converge faster without sacrificing the ability to navigate narrow parts of the geometry, thus avoiding stagnation by adapting the step size to the local geometry at each iteration. With this addition, the dependence of the QNGD's performance on the chosen step size is mitigated. Although this method introduces new hyperparameters to set, these are less problematic to choose than the step size.
This adaptive quantum natural gradient descent~(AQNGD) algorithm is detailed in Appendix~\ref{app:AQNGD}.

Notably, the Fubini-Study metric does not mitigate barren plateaus (BP). Because BPs are defined as the exponential decay of the gradient amplitude, merely rescaling the gradient with a metric cannot recover the lost signal. In fact, the gradient amplitude is dictated by the observable's locality, circuit depth, and initial state complexity~\cite{larocca2025barren}. Additionally, in the case of noise-induced barren plateaus the information is completely lost, hence the metric is completely ineffective.

\subsection{Euclidean Ansatz}
\label{sec:ansatz}
\begin{figure*}
\centering
    \begin{subfigure}[c]{0.25\textwidth}
    \centering
    \includegraphics{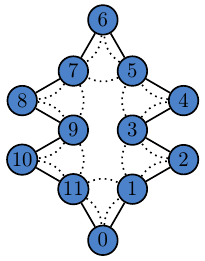}
    \caption{}
    \label{fig:kagome-star-map}
    \end{subfigure}%
    \hfill
    \begin{subfigure}[c]{0.72\textwidth}
    \centering
    \includegraphics[width=\linewidth]{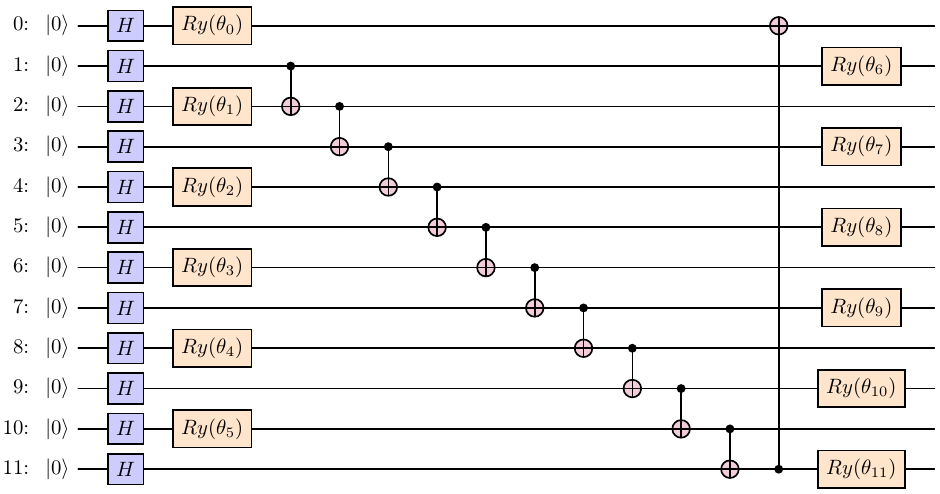}
    \caption{}
    \label{fig:star-ansatz-circuit}
    \end{subfigure}
\caption{(a) Hardware qubit connection is shown by solid lines, while Heisenberg interactions are shown in dotted lines. (b) The ansatz circuit used to prepare a trial wavefunction for the kagome star case. This ansatz is hardware compatible: It is linearly entangling, restricted to the hardware connectivity, and it is low depth with only one entangling layer and 12 parameterized rotation gates.}
\label{fig:star-ansatz}
\end{figure*}
Ideally, the ansatz for a specific physical problem should satisfy three conditions: (i) \textit{Expressivity}, i.e., the ability of the ansatz to prepare the system's ground state. (ii) \textit{Hardware efficiency}, ensuring that meaningful information survives the inherent QPU noise. (iii) \textit{Trainability}, meaning the parameter space is well-behaved so that the classical optimizer can efficiently converge to the optimal parameters.
The best physically inspired ans\"atze designs for the kagome lattice so far have been based on the Hamiltonian variational ansatz (HVA), which can be viewed as a digital approximation of adiabatic evolution. \citet{Park2024} proved that, in the absence of incoherent noise and under specific constraints, the HVA does not suffer from barren plateaus. \citet{Bosse2022} optimized the number of layers required for the HVA on QPUs with a square-lattice architecture. \citet{Kattemolle2022} proposed an HVA-inspired symmetry-preserving ansatz using dense layers of parametrized Heisenberg interaction gates starting from a dimer state, and further predicted the error rate and circuit depth required to achieve a given fidelity for arbitrary system sizes. A more hardware-friendly ansatz targeting valence-bond states (VBS) composed of dimer bonds was introduced by \citet{Sajjan2023}.
In this work, our goal is to
develop an expressive, trainable, hardware-efficient ansatz that can be implemented on current QPUs without relying on heavy problem-specific reductions, but only hardware constraints. To achieve this, we construct an ansatz based on a simple pattern of parametrized single qubit rotations combined with entangling gates between physically connected qubits to avoid SWAP gates. Our ansatz, shown in Fig.~\ref{fig:star-ansatz}, is inspired by the well-known \emph{RealAmplitudes} ansatz, but reduces the number of $R_y$ gates. Specifically, we found that placing $R_y$ rotations on every qubit both before and after the entangling layer was unnecessary. Instead, we alternate their placement, splitting them between the two sides of the entangling layer so that each qubit carries only a single $R_y$ rotation. This choice serves to reduce the number of parameters, making the ansatz more trainable, while also mitigating some of the coherent noise coming from the entangling layer through the parametrized gates after it.
Although the choice of a ladder $\operatorname{CNOT}$ structure increases depth, a more compact brick-wall structure will increase coherent crosstalk noise~\cite{Murali2020,mckay2023}.\\
Beyond this structural simplification, a central feature of our ansatz is that it naturally has a parameter space with a Euclidean geometry. By analytically evaluating the Fubini–Study metric (see Appendix~\ref{app:block-diagonal-fubiny-study}), we show that it is diagonal and constant for our construction. This means that the parameters vary independently (i.e., no over-parameterization), the parameter space is free of singularities, and the gradients obtained via the parameter-shift rule coincide with the true natural gradients. In other words, our ansatz is not only hardware efficient, but also intrinsically trainable. Since the Fubini-Study metric is diagonal and constant, the \textit{number condition} (i.e., the ratio of the largest eigenvalue to the smallest eigenvalue) is minimum. This implies that the metric is stable as recently shown by~\citet{Haddou2025}. In this setting, AQNGD is equivalent to just Adaptive Gradient Descent and will be referred to as AGD for the rest of the paper.
Altogether, this ansatz fulfills the three conditions outlined above: It is expressive, hardware efficient, and intrinsically trainable, thanks to its Euclidean parameter space. Our results confirm its expressivity, as it successfully prepares the ground state and reproduces physical observables such as the static spin structure factor $S(\boldsymbol{q})$. Looking forward, this construction opens the way for further investigations into scaling this family of Euclidean hardware-efficient ans\"atze to larger kagome lattices and, more broadly, to other frustrated quantum magnets.
\section{Experimental Results}
\label{sec:results}
In this section, we expose the variational quantum eigensolving for the ground state experiments on quantum devices. The purpose is to test the performance of optimizers and quantum error mitigation techniques in real situations. First, we show the experiments on three qubits for a system of three spin-$1/2$ sites. Then, we show experiments for $12$~Heisenberg interacting spin-$1/2$ sites connected in one kagome star.
\subsection{Antiferromagnetic Heisenberg interacting $1/2$~spins triangle}
\label{sec:KAHM-triangle}
\begin{figure}[t]
\centering
\begin{subfigure}[c]{0.28\linewidth}
    \centering
    \includegraphics{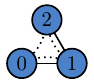}
    \caption{}
\end{subfigure}%
\hfill
\begin{subfigure}[c]{0.68\linewidth}
    \centering
    \includegraphics[width=\linewidth]{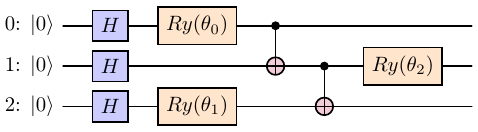}
    \caption{}
\end{subfigure}
\caption{(a) Hardware qubit connection is shown by solid lines, while Heisenberg interactions are shown in dotted lines. (b) The ansatz circuit used as trial wavefunction in the triangle case. This ansatz is hardware compatible: First, it is linearly entangling, following the native hardware connectivity. Second, it is low depth with only one entangling layer and three parameterized rotation gates.}
\label{fig:triangle-ansatz}
\end{figure}
\begin{figure}[t]
	\centering
		\centering
        \includegraphics[width=\linewidth]{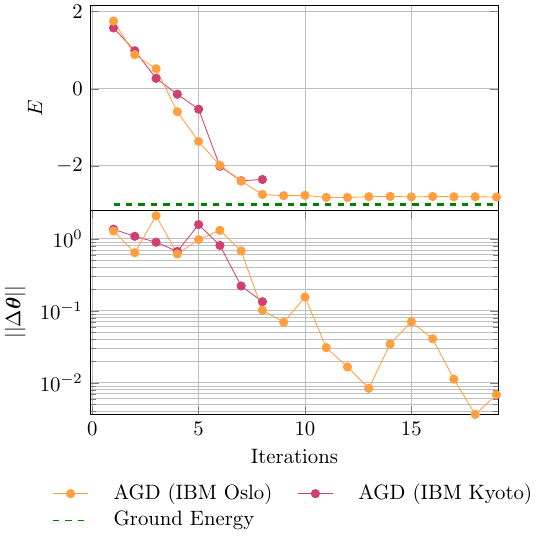}
    \caption{The upper plot displays the VQE convergence graph by iterations for the Heisenberg model on a triangle lattice, run on the IBM Kyoto 127-qubit and IBM Oslo 7-qubit QPUs, using the AGD optimization method. The REM and ZNE quantum error-mitigation techniques are not applied during the VQE optimization. The lower plot represents the step-size magnitudes $\lVert \Delta \boldsymbol{\theta}^\Delta_i \rVert = \lVert \boldsymbol{\theta}^\Delta_{i+1} - \boldsymbol{\theta}^\Delta_i \rVert = (\beta / 2^{k_i}) \lVert \nabla \boldsymbol{\theta}^\Delta_i \rVert$, which are determined by the product of the gradient-vector norm and the learning rate $\beta / 2^{k}$ set by the Armijo adaptive number $k$.}
    \label{fig:triangle-kyoto}
\end{figure}
Antiferromagnetic Heisenberg interacting spins in one triangle have three degenerate ground states whose energy value is $E_{\text{GS}}^{\Delta}=-3$. We show in the upper plot of Fig.~\ref{fig:triangle-kyoto}, the variational energy as evaluated on IBM Oslo (7 qubits) and IBM Kyoto (127 qubits) using the AGD optimization that is able to smoothly converge on three linearly connected qubits. The VQE algorithms have been implemented for nineteen iterations on IBM Oslo. Moreover, we performed an additional experiment on a newer quantum hardware platform, IBM Kyoto, using a limited number of iterations sufficient to achieve satisfactory convergence. The native gates of both QPUs are detailed in Appendix~\ref{app:SI}. The two plots show a very close behavior, even though the initial ansatz parameters vector $\boldsymbol{\theta}^\Delta_\text{init}$, in the same index order as Fig.~\ref{fig:triangle-ansatz}, is randomly picked. Namely, $$\boldsymbol{\theta}^\Delta_{\text{init}} = \left(0.2427\pi,\quad 0.2510\pi,\quad 0.1233\pi\right)$$
are given for IBM Kyoto experiment, while $$\boldsymbol{\theta}^\Delta_{\text{init}} = \left(0.2914\pi,\quad 0.2812\pi,\quad 0.1266\pi\right)$$
for IBM Oslo experiment.
It is worth recalling that the Fubini-Study metric is constant due to the structure of the quantum circuit, such that $g = 0.25 \cdot{I}_{3\times 3}$ as detailed in Appendix~\ref{app:block-diagonal-fubiny-study}.
The step size is adapted using line search backtracking as described in Sec.~\ref{sec:aqngd}. The number of executions $k$ required to find the best step size in each iteration is plotted in Fig.~\ref{fig:triangle-aqngd-combined} . The lower graph in Fig.~\ref{fig:triangle-kyoto} shows clearly the working mechanism of the learning rate adaptation due to the Armijo's rule fulfillment. In the beginning, the initial large step-size is enough to start exploration, then, it tends to slow down reciprocally with the increasing Armijo's adaptive number $k$ value to make the step size adaptive to the cost function landscape.
We notice that the step size continues decreasing in IBM Oslo experiment due to diminishing quantum natural gradient norm and the flexible learning rate. 
Overall, the convergence of the AGD optimized VQE energy curves in Fig.~\ref{fig:triangle-kyoto} is remarkable. Even though the experiment is small and the quantum circuit is shallow, it showcases the low noise rate of the real quantum devices and the trainability of the ansatz using quantum natural gradient descent and Armijo's adaptive method.\\
We note that the final trained ansatz parameters vector $\boldsymbol{\theta}^\Delta_\text{final}$ is given by
\begin{equation}
    \label{eq:params-triangle}
\boldsymbol{\theta}^\Delta_{\text{final}}=
    (0.2306\pi,\quad 1.5005\pi,\quad 1.0031\pi)
\end{equation}
for IBM Kyoto, in the same index order as in Fig.~\ref{fig:triangle-ansatz}, and
\begin{equation}
\boldsymbol{\theta}^\Delta_{\text{final}}=
    (0.3111\pi,\quad 1.5240\pi,\quad 1.0144\pi)
\end{equation}
for IBM Oslo experiment,
which are very close to the exact parameter values $\theta^\Delta_1=3\pi/2, \theta^\Delta_2=\pi$, while $\theta^\Delta_0$ is free to have any value. Typically, when $\theta_0^\Delta = \pi/2$, we can reproduce one of the theoretical ground states $\vert\psi^\Delta_{\text{GS}} \rangle$, which is:
\begin{equation}
    \ket{+} \otimes \frac{1}{\sqrt{2}} \left( \ket{0}\otimes\ket{1} - \ket{1}\otimes\ket{0} \right) = \ket{+} \otimes [1, 2].
\end{equation}
That is nothing but a dimer state, with the fidelity value $\vert\langle \psi_{\text{SVS}}(\boldsymbol{\theta}^\Delta_{\text{final}})\vert\psi^\Delta_{\text{GS}} \rangle\vert^2 = 99.89\%$ on the state-vector simulation~(SVS) and an error of $\Delta E = E_{\text{SVS}}(\boldsymbol{\theta}^\Delta_{\text{final}}) - E^\Delta_{\text{GS}}= 9\cdot 10^{-5}$ using the VQE-optimized parameters in Eq.~\eqref{eq:params-triangle} from IBM Kyoto.

\begin{figure}
    \centering
    \includegraphics[width=\linewidth]{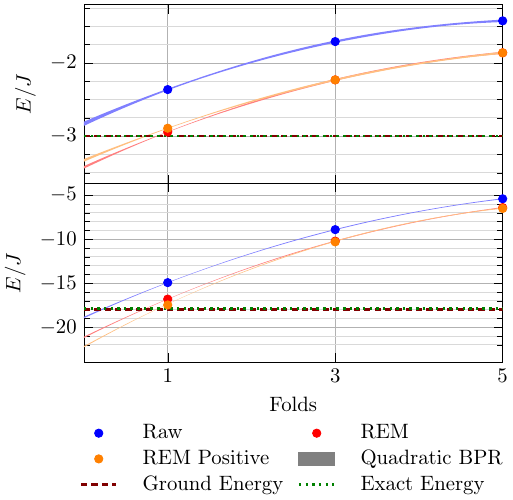}
    \caption{Benchmarking quantum error mitigation using ZNE and REM techniques as described in Sec.~\ref{sec:zne} and \ref{sec:rem}, where Bayesian polynomial regression (BPR) is used. Readout and zero-noise error mitigation evaluated on the IBM Kyoto of the ground-state energy measurement of antiferromagnetic spin-$1/2$ Heisenberg model on one triangle (top), and on one kagome star in IBM Torino (bottom).}
    \label{fig:zne-rem}
\end{figure}

Nevertheless, the exact parameters are not enough if the quantum state cannot be experimentally characterized. To enhance the accuracy of the energy measurement of the last state, error mitigation (EM) techniques are explored. We recall that EM was not used during the VQE procedure. The values of the energy prior to and subsequent to EM implemented on the IBM Kyoto machine are plotted in the upper plot of Fig.~\ref{fig:zne-rem}. The unmitigated value is found to be $-2.3634 \pm 0.0110$ in the normalized units, while the readout error mitigation using the full calibration matrix gives $-2.9423 \pm 0.0034$ with straightforward REM, and $-2.8917 \pm 0.0046$ with positive probabilities preserving REM, both showing better accuracy. Applying ZNE error mitigation on top of REM using quadratic extrapolation exceeds the ground energy as shown in Fig.~\ref{fig:zne-rem}. On the other hand, applying ZNE on the raw energy values in the first folding points yields the values $-2.8306 \pm 0.0286$ for quadratic extrapolation, which is the minimal polynomial that fits the three points. It is observed that the REM and the ZNE work better separately than in combination. These results will be discussed in Sec.~\ref{sec:discussion}.

Remarkably, the detection of local dimers can be enhanced with the REM method as shown in Tab.~\ref{tab:triangle-correlators}. The error mitigation makes the gap clear between spin correlations of different pairs when the dimer bond is present or absent. This observation fits qualitatively the theoretical expectation from the exact ground state, while quantitative closeness is enhanced through REM.
\begin{table}[t]
    \centering
    \begin{tabular}{cccccc}
    \hline\hline
    \multirow{3}*{Edges $(i,j)$} & \multicolumn{5}{c}{Correlation $\left\langle\boldsymbol{S}_i \cdot \boldsymbol{S}_j\right\rangle$} \\
    \cline{2-6}
    & \multicolumn{2}{c}{IBM Oslo} & \multicolumn{2}{c}{IBM Kyoto} & \multirow{2}*{Exact}\\
    \cline{2-5}
    & Raw & REM & Raw & REM &\\
    \hline
    (0,1) & $-0.0756$ & $-0.0636$ & $0.1476$ & $0.0701$ & $0$\\
    (1,2) & $-2.7828$ & $-2.9902$ & $-1.7740$ & $-2.8604$ & $-3$\\
    (2,0) & $0.0472$ & $0.0440$ & $-0.7288$ & $-0.1434$ & $0$\\
    \hline\hline
    \end{tabular}
    \caption{Correlation terms between spins of each edge pair in the triangle.}
    \label{tab:triangle-correlators}
\end{table}
\begin{table*}[t]
    \centering
    \renewcommand{\arraystretch}{1.4}
    \begin{tabular}{llll}
    \hline\hline
\multirow{2}*{Method}& \multicolumn{3}{c}{Energy Value $\pm \epsilon$ ($E/J$)}\\
        \cline{2-4}
                         & IBM Torino (a)        & IBM Algiers (b)       & IBM Torino (c)\\
        \hline
        Unmitigated      & $-14.9316 \pm 0.0265$ & $-10.9748 \pm 0.0343$ & $-14.6346 \pm 0.0260$\\
        \hline
        REM              & $-16.8028 \pm 0.0195$ & $-12.4031 \pm 0.0306$ & $-16.2172 \pm 0.0199$\\
        REM + Positivity & $-17.3675 \pm 0.0123$ & $-12.1712 \pm 0.0314$ & $-16.0097 \pm 0.0202$\\
        \hline
        Polynomial deg=1 ZNE
                         & $-17.2644 \pm 0.0355$ & $-12.8618 \pm 0.0441$ & $-16.3731 \pm 0.0348$\\
        Polynomial deg=2 ZNE
                         & $-18.8826 \pm 0.0790$ & $-18.0060 \pm 0.0922$ & $-17.9740 \pm 0.0780$\\
        \hline
        REM + Polynomial deg=2 ZNE
                         & $-21.1330 \pm 0.0705$ & $-20.3412 \pm 0.0877$ & $-19.8208 \pm 0.0705$\\
        REM + Positivity + Polynomial deg=2 ZNE
                         & $-22.2032 \pm 0.0640$ & $-19.9411 \pm 0.0888$ & $-19.4483 \pm 0.0707$\\
        \hline
        Exact Energy     & $-17.8533$            & $-17.9998$            & $-17.9998$\\
\hline
        Ground Energy    & \multicolumn{3}{c}{$-18.00000$}\\
        \hline\hline
    \end{tabular}
    \caption{(a) Results for the AGD driven VQE for one kagome star performed on IBM Torino QPU as shown in Fig.~\ref{fig:star-torino} and mitigated using REM and ZNE as shown in Fig.~\ref{fig:zne-rem}. (b) IBM Algiers ground state measurement and mitigation for one kagome star. (c) IBM Torino ground-state energy measurement and mitigation for one kagome star. The parameters of (b) and (c) experiments are obtained from classically simulated VQE.}
    \label{tab:star-QEM}
    \end{table*}
\subsection{Antiferromagnetic Heisenberg interacting $1/2$~spins kagome star}
\label{sec:KAHM-star}
The antiferromagnetic Heisenberg model of $1/2$~spins on a kagome star has two possible dimer ground states,

\begin{align}
    \ket{\psi^1_{\text{GS}}} &= [1, 2] \otimes [3, 4] \otimes \cdots \otimes [11, 0].
\end{align}
which is the unique dimer state that is expected from the ansatz in Fig.~\ref{fig:star-ansatz}. The other state should be
    \begin{align*}
        \ket{\psi_\text{GS}^2} =[0, 1] \otimes [2, 3] \otimes \cdots \otimes [10, 11],
    \end{align*}
where the qubits are labeled as shown in Fig.~\ref{fig:star-ansatz}.
Both states correspond to the ground energy $E_{\text{GS}}=-18$, since the kagome star is constructed by simply connected triangles as discussed in Sec.~\ref{sec:introduction}. The ansatz depicted in the quantum circuit in Fig.~\ref{fig:star-ansatz} targets only the first state. Although this introduces a physical bias in the ansatz, it serves as a benchmark to the quantum device, allowing us to assess the performance of a quantum computer on such problems in the NISQ regime. To assess the benefits of AGD's adaptivity and its exact gradient, we compare it with SPSA. \\

\begin{figure}[t]
	\centering
    \includegraphics[width=\linewidth]{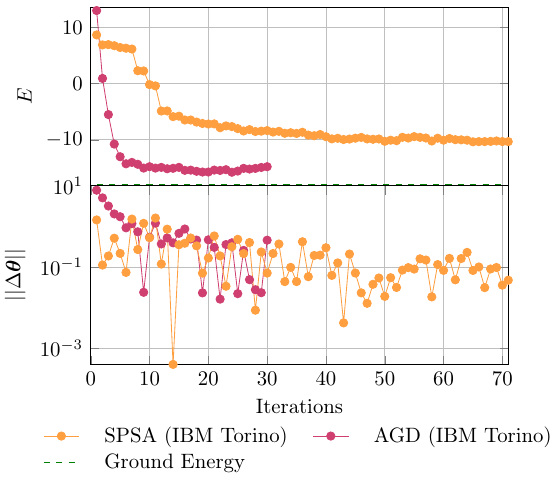}
    \caption{The upper plot shows the VQE convergence graph per iteration as illustrated in terms of energy variation for the Heisenberg model on a single kagome star lattice run on IBM Torino 133 qubits QPU with the AGD and SPSA optimizers. Notice that AGD requires more quantum circuit executions per iteration than SPSA to evaluate the gradient (one execution for each the 12 parameters) and to adapt the learning rate, which is reflected in the number of executions $k$ in Fig.~\ref{fig:star-aqngd-combined}. However, most of those executions are independent of each other and hence parallelizable.
    The lower plot represents the step-size magnitudes $\lVert \Delta \boldsymbol{\theta}_i \rVert = \lVert \boldsymbol{\theta}_{i+1} - \boldsymbol{\theta}_i \rVert$, which are determined by the product of the gradient-vector norm and the learning rate $\beta / 2^{k}$ set by the Armijo adaptive number $k$ in the AGD case.}
    \label{fig:star-torino}
\end{figure}

The VQE convergence graph is depicted in Fig.~\ref{fig:star-torino} for both SPSA and AGD optimizers. The final components of the trained ansatz parameters vector $\boldsymbol{\theta}_\text{final}$, in the same index order as Fig.~\ref{fig:star-ansatz}, for the AGD optimization experiment are given, up to a global phase, by
\begin{equation}
\label{eq:params-star}
    \left(
    \begin{array}{c|c}
    \theta_0 =-0.3962 \pi & \theta_6 = +1.0056 \pi\\
    \theta_1 =-0.5324 \pi & \theta_7 =+1.0190 \pi\\
    \theta_2 =+1.4827 \pi & \theta_8 =+0.9836 \pi\\
    \theta_3 =-0.4658 \pi & \theta_9 =-0.9768 \pi\\
    \theta_4 =-0.4879 \pi & \theta_{10} =+1.0007 \pi\\
    \theta_5 =+1.4993 \pi & \theta_{11} =-0.9830 \pi
    \end{array}
    \right)
    \approx
    \left(
    \begin{array}{c|c}
    -\pi/2 & \pi\\
    -\pi/2 & \pi\\
    -\pi/2 & \pi\\
    -\pi/2 & \pi\\
    -\pi/2 & \pi\\
    -\pi/2 & \pi
    \end{array}
    \right),
\end{equation}
where the right side defines the converged exact ansatz parameters of the ground state. Namely, according to the ansatz in Fig.~\ref{fig:star-ansatz}, the parameterized state $\ket{\psi(\boldsymbol{\theta}_\text{final})}$ should yield the following state $\ket{\psi^1_{\text{GS}}}$.

\begin{figure}[t]
	\centering
	\includegraphics[width=\linewidth]{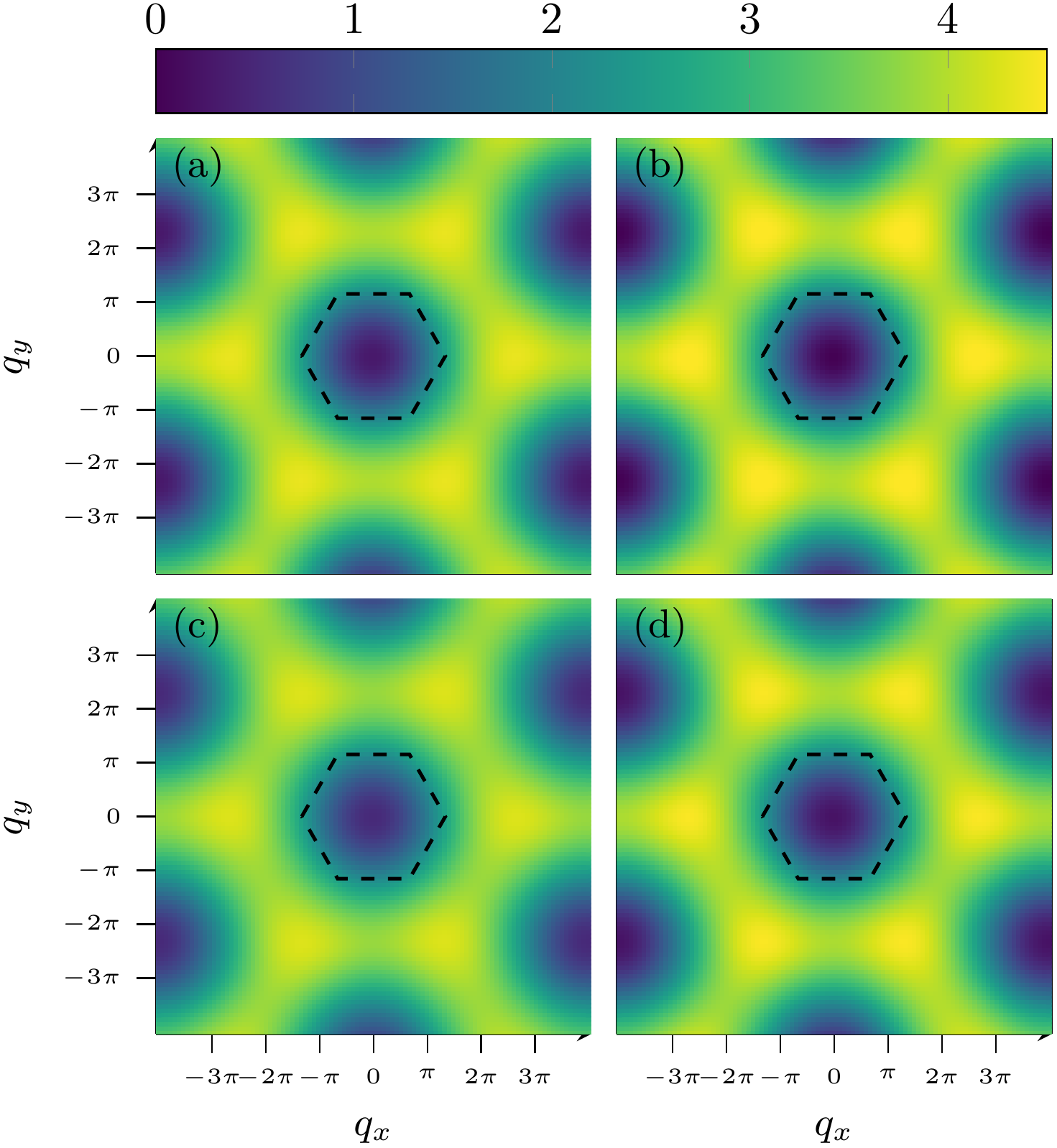}
    \caption{Spin-spin correlation static spin structure factor $S(\boldsymbol{q})$ (the Fourier transform of the equal-time spin-spin correlation function) as computed from the measurement results of noisy simulation of IBM Torino device (a), ideal simulator with shot noise (b), real IBM Torino QPU without error mitigation (c), and with REM (d). The dashed hexagon draws the border of the Brillouin zone. The suppression at zero momentum indicates the absence of ferromagnetic order.}
    \label{fig:star-torino-Sq}
\end{figure}

Using the parameters in Eq.~\eqref{eq:params-star}, we get the fidelity value $\vert\langle\psi_{\text{SVS}}(\boldsymbol{\theta}_{\text{final}})\vert\psi_{\text{GS}}\rangle\vert^2=96.37\%$ and $\Delta E = E_{\text{SVS}}(\boldsymbol{\theta}_{\text{final}}) - E_{\text{GS}}= 0.1463$ in the state-vector simulator when the result is compared with the ground state and the ground energy, respectively.\\
To retrieve more accurate information from the quantum results, we exploit the REM and the ZNE techniques. First, we show in Fig.~\ref{fig:zne-rem}, the results of error mitigation techniques. The raw energy value is $-14.9316 \pm 0.0265$ which is around $3$ units away from the theoretical value. However, with REM only, we get $-16.8028 \pm 0.0195$ and $-17.3675 \pm 0.0123$ for straightforward REM and positivity preserving REM, respectively. Using ZNE only with Bayesian quadratic extrapolation gives $-18.8826 \pm 0.0790$.
On the other hand, the combination of REM and ZNE undershoots beyond $-19$ as detailed in Tab.~\ref{tab:star-QEM}. These results are consistent with the similar behavior of REM and ZNE in the triangle case in Sec.~\ref{sec:KAHM-triangle}.\\
For more insight, we run two different error mitigation experiments on the IBM Torino and IBM Algiers quantum computers. The results are detailed in Tab.~\ref{tab:star-QEM}. In these experiments, we use parameters from a classically simulated VQE.
Even though REM is not consistently well performing as in IBM Algiers, it provides an upper bound especially when the positivity of sampling counts is retained.
On the other hand, we observe again the matching between the ZNE with Bayesian quadratic extrapolation and the theoretical value. This observation is also consistent with the ZNE experiment in the triangle case.\\
Noticeably, the spin correlation structure can be effectively characterized even with the noisy backend, as shown in Fig.~\ref{fig:star-torino-Sq}. Here, the static spin structure factor is computed based on the measurement of each spin-to-spin correlation, and which is defined as:
\begin{align}
    S(\boldsymbol{q}) &=
    \frac{1}{N} \sum_{ij} e^{i\boldsymbol{q}\cdot (\boldsymbol{r}_i-\boldsymbol{r}_j)} \left\langle \boldsymbol{S}_i \cdot \boldsymbol{S}_j \right\rangle,
\end{align}
where the summation goes over all spins in sites $\boldsymbol{r}_i$ for $i=1,\cdots,N$, and $\boldsymbol{q}$ is the momentum vector in the reciprocal lattice.
The static structure factor reveals the signature of the magnetization order of the system. This is the Fourier transform of the equal-time spin-spin correlation function, not merely the Fourier transform of the lattice geometry, and thus reflects the correlation patterns between spins in the quantum state. The comparison between the exact and experimental spin structure factor shows that errors in the individual spin correlations are not enough to erase the correlation order signature. Moreover, we exploit REM method to enhance the landscape of the spin structure factor.
The Pearson correlation coefficient indicates similarities of $99.97\%$, $99.97\%$, and $99.94\%$ with the exact
$S(\boldsymbol{q})$ for the simulated noisy results, the QPU results, and the REM-mitigated QPU results, respectively. The corresponding mean-square errors are $0.013$, $0.042$, and $0.007$, in the same order.
Hence, REM balances between accuracy and qualitative pattern similarity.
\section{Discussion}
\label{sec:discussion}
\begin{figure*}[t]
    \centering
    \begin{subfigure}{0.48\textwidth}
    \centering
        \includegraphics[width=\linewidth]{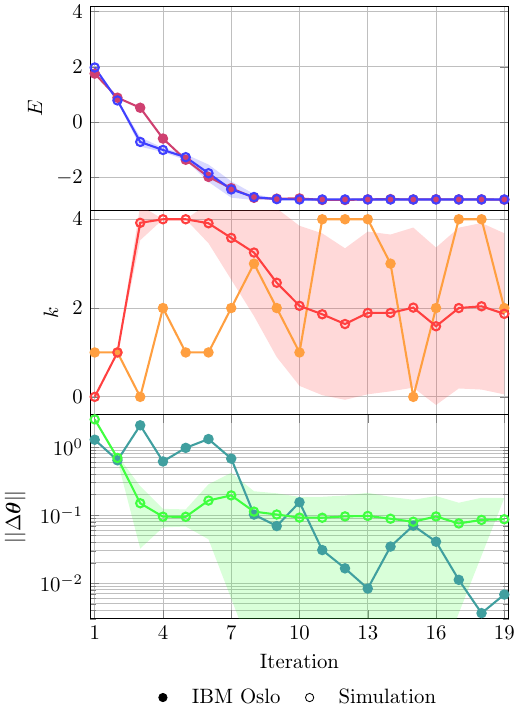}
        \caption{Triangle - IBM Oslo vs Simulation}
        \label{fig:triangle-aqngd-combined}
    \end{subfigure}
    \begin{subfigure}{0.48\textwidth}
    \centering
        \includegraphics[width=\linewidth]{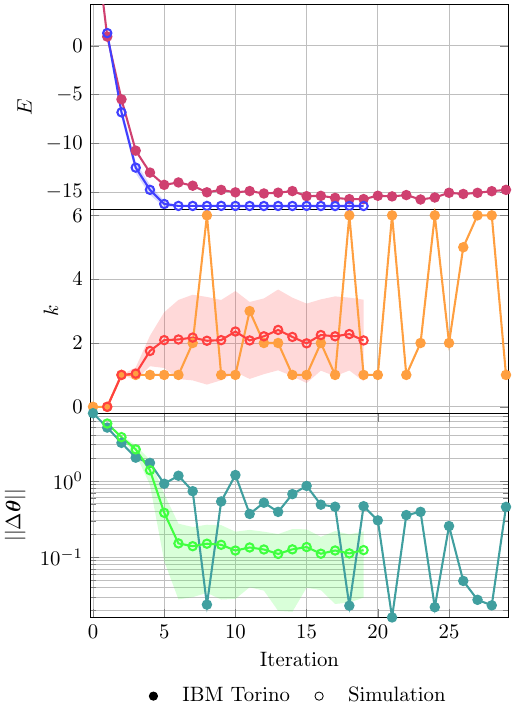}
        \caption{Star - IBM Torino vs Simulation}
        \label{fig:star-aqngd-combined}
    \end{subfigure}
    \caption{AGD optimization results for the Heisenberg model on the triangle lattice comparing IBM Oslo QPU and simulation of the VQE experiment on IBM Oslo noise model over 100 runs~(a); and on one kagome star lattice comparing IBM Torino QPU and simulation of the same VQE experiment on IBM Torino noise model over 100 runs, starting from the same initial parameters~(b). The step norm $\vert\vert \Delta \boldsymbol{\theta} \vert\vert = \beta / 2^k \vert\vert \nabla \boldsymbol{\theta}\vert\vert$ in both cases.}
    \label{fig:AQNGD-exp-results}
\end{figure*}
\begin{figure}[h]
    \centering
        \includegraphics[width=\linewidth]{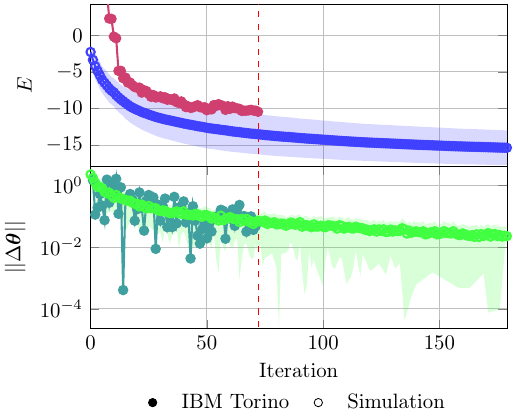}
    \caption{SPSA optimization results for the Heisenberg model on one star kagome lattice comparing IBM Torino QPU and simulation of the same VQE experiment on IBM Torino noise model over 100 runs, starting from the same initial parameters~(a). The step norm $\vert\vert \Delta \boldsymbol{\theta}_i \vert\vert = \lVert \boldsymbol{\theta}_{i+1} - \boldsymbol{\theta}_i\rVert$.}
    \label{fig:star-spsa-best-k-SM-full}
\end{figure}
Here, we will discuss the ability of AGD to prepare the ground state, the need to measure noise resilient properties such as the static structure factor, and we discuss further the enhanced energy values using EM.\\
In Fig.~\ref{fig:star-torino}, we can clearly see the superior optimization performance of AGD over SPSA in real, noise-affected, quantum devices. AGD requires fewer iterations to converge compared to the inherently noise-adapted SPSA method. Even though the number of circuit executions is larger than what is required by an SPSA implementation, scaling linearly in the number of parameters, AGD can be boosted by distributed computation in each iteration and is guaranteed to converge and thus more scalable than zeroth-order stochastic optimization methods~\cite{Stokes2020, Atif2023}.
To assess the role of adaptation in the AGD performance, it is sufficient to check the profile of the step-size $\lVert \Delta \boldsymbol{\theta}\rVert$ during the optimization.
The lower plot of Fig.~\ref{fig:star-torino} showcases the comparison between the magnitudes of the step-sizes taken by AGD and SPSA. First, we notice that the profile of Armijo's adapted step-size shows qualitatively similar evolution pattern as SPSA with exponentially decaying learning rate~\cite{Nacer2025}. Second, the Armijo rule allows to take arbitrary larger initial steps to converge faster as shown in the same figure, while adapting the learning rate to fine tune when needed. These observations are supported by simulation experiments in Fig.~\ref{fig:AQNGD-exp-results} and Fig.~\ref{fig:star-spsa-best-k-SM-full}.\\
In addition, as shown in Sec.~\ref{sec:KAHM-triangle} and Sec.~\ref{sec:KAHM-star}, despite the significant noise-induced gap in the energy of the ground state, the optimized parameters show impressive accordance with the exact ground state parameters without any reliance on error mitigation techniques. This observation can be quantified by evaluating the fidelity and energy values using the state-vector simulation of the ans\"atze. Namely, the state-vector fidelities are $99.89\%$ in the case of one triangle, and $96.37\%$ in the case of one kagome star, while energy gaps are $\Delta E = 9\cdot 10^{-5}$ and $\Delta E = 0.1463$ for the triangle and the kagome star, respectively.\\
Although this observation highlights the promising capabilities of the VQE protocol to prepare accurate states despite errors~\cite{Farshud2025,Nacer2025}, it becomes impractical and not scalable if such accurate states are inaccessible by quantum computing means. Our results show that characterizing the dimer state using spin correlation and spin structure factor is possible even without error mitigation. Moreover, it is possible to enhance the state's spin structure resolution through error mitigation means such as REM as shown in Fig.~\ref{fig:star-torino-Sq}. This result is in agreement with the recent work of \citet{Lively2024} who showed that the prediction of the energy derivative, two-site spin correlation functions, and the fidelity susceptibility from VQE results can be accurate even without error mitigation. For instance, local observables are less sensitive to noise than non-local ones~\cite{Kim2023}.

\begin{figure}[t]
	\centering
	\includegraphics[width=\linewidth]{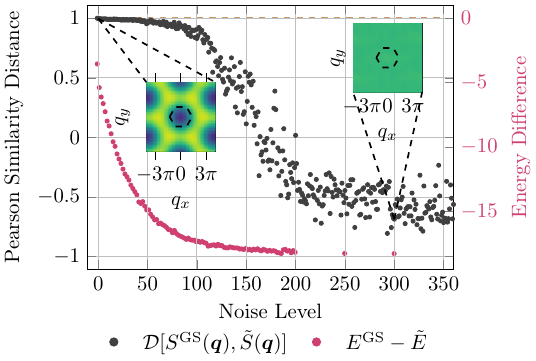}
    \caption{Pearson similarity distance $\mathcal{D}[S^\text{GS}(\boldsymbol{q}), \tilde{S}(\boldsymbol{q})]$ between the spin–spin correlation Static Structure Factor (SSF) of the prepared state under noise ($\tilde{S}(\boldsymbol{q})$) and the ideal spin-spin correlation SSF ($S^\text{GS}(\boldsymbol{q})$) as a function of noise level. The noise strength is introduced by controlled folding of arbitrary one- and two-qubit gates, in this case, the sequence ($\text{H}\cdot \text{CX}\cdot \text{CX}\cdot \text{H}$).
    For comparison, the same plot includes the energy deviation, defined as $E^\text{GS} - \tilde{E}$ where $E^{GS}=-18$ is the ground-state energy and $\tilde{E}$ is the energy of the same state under incoherent noise.}
    \label{fig:ssf-noise-resilience}
\end{figure}

One of the main results of this work is that the spin-spin correlation can be noise-resilient. Noise-resilience is in the sense that it is steady under some decoherence noise threshold. To demonstrate that, we conduct many simulation experiments where we measure the spin-spin SSF in different noise levels, as shown in Fig.~\ref{fig:ssf-noise-resilience}.
As the noise level increases, the SSF similarity error does not exhibit linear or polynomial scaling. Instead, it displays a step-like, sigmoid-shaped dependence, characterized by two quasi-plateau regimes separated by a sharp crossover region. In the low-noise regime, the SSF remains nearly constant and close to the ideal value, indicating robustness of the correlation structure. Beyond a critical noise threshold, the SSF rapidly transitions to a second plateau where correlations are substantially degraded. Although the transition is continuous, its sharp crossover resembles an approximate phase transition between two distinct correlation regimes. Notably, the width of the crossover coincides with the region where the SSF remains physically meaningful.
In contrast to the SSF behavior under noise, the energy exhibits a monotonic decay toward the mixed-state value as noise increases. This inverse trend highlights that while energy degrades gradually under noise, the SSF retains structural integrity up to a threshold and then undergoes a rapid qualitative change, emphasizing the nontrivial sensitivity of correlation-based observables to noise amplification.
We should direct the attention of the reader to the fact that the noise strength is arbitrary and unitless. It is introduced by controlled folding of one- and two-qubit gates following the sequence ($\text{H}\cdot \text{CX}\cdot \text{CX}\cdot \text{H}$), effectively amplifying the underlying hardware noise without altering the target unitary. It is important to say that such noise resilience is expected since the spin-spin correlation is featured by a local dimer formation energy gap that protects the SSF.\\
On the other hand, we should emphasize that a mere two-site descriptor is not sufficient to distinguish between arbitrary two pure quantum states. Namely, \citet{Bosse2022} have shown that spin-spin correlation structure can be very close to the ground state while the state fidelity is very low. Nevertheless, they showed also that VQE states can represent these local observables well even when they have not yet reached high global fidelity.

Finally, error mitigation (EM) techniques are of great importance in this context. The results of error mitigation implemented on the last ansatz obtained from the VQE for one kagome star are detailed in Tab.~\ref{tab:star-QEM}. Although the results are strongly stochastic, the accuracy is enhanced using different EM techniques. First of all, we notice that REM alone fails in the IBM Algiers' case to reach a significant accuracy despite its success in other cases. The combination of REM and ZNE overshoots in most cases. Impressively, ZNE with quadratic extrapolation obtained repeatedly the best accuracy across different experiments, as shown in Fig.~\ref{fig:zne-rem} and Tab.~\ref{tab:star-QEM}. The combination of REM and ZNE overshoots the ground energy, which is consistent with the results of~\citet{Kurita2023} that show that ZNE yields a better accuracy when combined with randomized compilation, to suppress coherent noise, in the absence of SPAM errors. In addition, ZNE without REM may go below the exact energy, but to a lesser extent. This is in part due to coherent noise channels shifting the measured energy values in a non-stochastic way,
which can potentially be captured by the ZNE fitting functions.
A careful study, currently the focus of an ongoing work by the authors, is necessary to provide a more solid explanation to this mechanism.
In the end, it is important to recall that REM does not break the variational principle as explained in Appendix~\ref{sec:rem}, while there is no known guarantee for this with ZNE.
\section{Conclusion}
\label{sec:conclusion}
In this work, we focused on the preparation of the ground state of kagome lattice fragments on real quantum devices, showing that this can be achieved with shallow circuits using the \textit{variational quantum eigensolver} (VQE). Remarkably, this holds even in the presence of significant noise and without applying error mitigation. While noise limits the quantitative accuracy of energy estimation, essential qualitative features—such as the spin correlation structure of the dimer state—remain accessible, owing to the energy gap that protects certain observables from noise effects.
We exploited the Fubini-Study metric as a tool to analyze the geometry of the ansatz and designed it to have a Euclidean parameter space (i.e., constant diagonal Fubini-Study metric), which allows for a smooth and easy optimization without facing singularities. Building upon this, we believe that this approach of carefully designing the ansatz's geometry holds significant potential. We posit that such a design methodology can be exploited to design problem-tailored ans{\"a}tze with analytically computable Fubini-Study metrics, which leads to expressive, hardware-efficient, and intrinsically trainable ans{\"a}tze at low costs. This suggests that this kind of ans{\"a}tze architecture could enable increasingly accurate and robust preparation of complex ground states for frustrated systems as quantum hardware continues to progress, including larger kagome lattices and quantum spin liquids.
We analyzed the performance of the \textit{adaptive gradient descent} (AGD), where backtracking search is used, in comparison with SPSA, showing that AGD achieves convergence in fewer iterations while maintaining competitive runtime, with implicit Euclidean parametrized space of ans{\"a}tze. We showed that the adaptive feature through backtracking search enables faster convergence and less dependency on the starting point.
Our results further reveal that error mitigation techniques can substantially improve the accuracy of observable estimates: \textit{readout error mitigation} (REM) preserves the variational principle, providing reliable upper bounds, while \textit{zero-noise extrapolation} (ZNE) does not respect this bound but often achieves the highest accuracy, suggesting potential for future studies. We also showed that certain observables, such as the static spin structure factor, are less sensitive to noise than the Hamiltonian, suggesting a good characterization of some of the ground state properties despite the challenges of the quantum hardware.\\
Finally, we highlight that solving device-native problems such as the one kagome star ground state on heavy-hex architectures not only enables realistic benchmarking but also paves the way for scaling to larger kagome lattices. Future work will focus on extending these methods to characterize the
ground state of larger quantum spin liquid systems.
\section*{Data Availability}
The QPU data and post-processing code for ZNE and REM that support the findings of this article are openly available
on Zenodo \cite{dataset2026} under DOI: 10.5281/zenodo.21264601.
\section*{Acknowledgments}
This document has been produced with the financial assistance of the European Union (Grant no. DCI-PANAF/2020/420-028), through the African Research Initiative for Scientific Excellence (ARISE), pilot programme. ARISE is implemented by the African Academy of Sciences with support from the European Commission and the African Union Commission. The contents of this document are the sole responsibility of the authors and can under no circumstances be regarded as reflecting the position of the European Union, the African Academy of Sciences, and the African Union Commission.
We are grateful to the Algerian Ministry of Higher Education and Scientific Research and DGRST for the financial support.
We thank the Quantum Collaborative for their support and access to IBM Quantum Resources. We acknowledge the use of IBM Quantum services for this work. The views expressed are those of the authors, and do not reflect the official policy or position of IBM or the IBM Quantum team.
The authors gratefully acknowledge fruitful discussions with Dr. Manas Sajjan.
\appendix
\section{Adaptive Quantum Natural Gradient Descent}
\label{app:AQNGD}
In the quantum natural gradient descent, the parameters are updated according to the gradient value, which is measured using the parameter-shift rule method that gives the Euclidean gradient, multiplied by the inverse of the Fubini-Study block diagonal metric, which squeezes and stretches the gradient according to the geometry of the parameter space. The latter is a practical approximation of the full Fubini-Study metric that would require exhaustive operations and techniques that are near impossible to implement on NISQ devices~\cite{Stokes2020}. Another near-term alternative is the efficient protocol by \citet{gomez2025} for lowering the number of executed circuits from $O(\text{number of parameters})$ to $O(\text{number of layers})$. The learning rate is adapted using the Armijo rule~\cite{Armijo1966} as detailed in Algorithm~\ref{alg:backtracking}. This subroutine searches for the largest learning rate that verifies the Armijo condition.
However, AQNGD requires the careful tuning of the following hyperparameters that can affect the performance of optimization:
\begin{itemize}
  \renewcommand\labelitemi{-}
  \setlength\itemsep{0.2em}
  \setlength\parskip{0pt}
  \setlength\parsep{0pt}
\item $\alpha \in [0,1] $: the constant to tune the sensitivity of Armijo's line search principle. Lower values of $\alpha$ skew the rule towards the preference of higher step-sizes, and higher values skew it towards the preference of lower step-sizes. In this work, we used $\alpha=0.01$.
\item $\beta $: the maximum learning rate (step size) used by the algorithm. It is fixed in this work to $\beta=0.5$.
\item $k_m$: the maximum number of step-sizes searched by the line search algorithm to determine the best step-size. In this work, $k_m=6$.
\item $\epsilon$: the tolerance parameter used in the \texttt{PseudoINVERT} function which is available as a Numpy function~\cite{2020NumPy-Array}.
In our case, $\epsilon$ is set to $10^{-15}$.
\end{itemize}
\begin{algorithm}[H]
\caption{ApplyArmijoRule.\label{alg:backtracking}}
\begin{algorithmic}[1]
	\Require $f$: Cost function, $\alpha$, $\beta$, $\boldsymbol{\theta}_i$
	\Ensure $k$
	\State Assert $\alpha \in [0, 1]$, $\beta > 0$
	\State $k \gets 0$
	\State $\boldsymbol{\theta}_{i+1} \gets \boldsymbol{\theta_i}-\frac{\beta}{2^k}\nabla f(\boldsymbol{\theta_i})$
	\State $\text{ArmijoCondition} \gets f(\boldsymbol{\theta}_{i}) - f(\boldsymbol{\theta}_{i+1}) \geq \alpha \frac{\beta}{2^k} \vert\vert \nabla f(\boldsymbol{\theta}_{i}) \vert\vert^2$
	\While{not ArmijoCondition}
	\State $k \gets k+1$
	\State $\boldsymbol{\theta}_{i+1} \gets \boldsymbol{\theta_i}-\frac{\beta}{2^k}\nabla f(\boldsymbol{\theta_i})$
	\State $\text{ArmijoCondition} \gets f(\boldsymbol{\theta}_{i}) - f(\boldsymbol{\theta}_{i+1}) \geq \alpha \frac{\beta}{2^k} \vert\vert \nabla f(\boldsymbol{\theta}_{i}) \vert\vert^2$
	\EndWhile
	\end{algorithmic}
\end{algorithm}
\begin{algorithm}[H]
\caption{AQNGD Algorithm : It follows typically the QNGD algorithm, but it is featured by an additional step to adapt the step size rate $\lambda_i$ using the line search backtracking that involves the optimization of the Armijo-Goldstein rule as detailed in Algorithm~\ref{alg:backtracking}.\label{algAQNGD}}
\begin{algorithmic}[1]
	\Require $f$ : Cost function, $\alpha$, $\beta$, $\boldsymbol{\theta}_0$, $\epsilon$
	\Ensure $f(\boldsymbol{\theta}_{i})$
	\State Assert $\alpha \in [0, 1]$, $\beta > 0$, $\epsilon > 0$
	\State $i \gets 0$
	\State $\boldsymbol{\theta}_i \gets \boldsymbol{\theta}_0$
	\Repeat
	\State $\nabla_{\mathrm{Euclid}}f(\boldsymbol{\theta}_i) \gets \mathrm{ParameterShift}(\boldsymbol{\theta}_i)$;
	\State $g(\boldsymbol{\theta}_i) \gets \mathrm{FubiniStudyMetric}(\boldsymbol{\theta}_i)$
	\State $g^{-1}(\boldsymbol{\theta}_i) \gets \mathrm{PseudoInvert}(g(\boldsymbol{\theta}_i),\epsilon)$
	\State $\nabla f(\boldsymbol{\theta}_i) \gets g^{-1}(\boldsymbol{\theta}_i) \nabla_{\mathrm{Euclid}}f(\boldsymbol{\theta}_i)$
	\State $k \gets \mathrm{ApplyArmijoRule}(\boldsymbol{\theta}_i, \alpha, \beta)$
	\State $\lambda_i \gets \beta/{2^k}$
	\State $\boldsymbol{\theta}_{i} \gets \boldsymbol{\theta}_{i} - \lambda_i \nabla f(\boldsymbol{\theta}_i)$
	\State $i \gets i+1$
	\Until $f(\boldsymbol{\theta}_{i-1}) - f(\boldsymbol{\theta}_{i}) < \epsilon$
	\end{algorithmic}
\end{algorithm}
\section{Heisenberg Hamiltonian Properties}
\label{app:heisenberg-hamiltonian-properties}
The Heisenberg Hamiltonian is defined as:
\begin{align}
    H = \sum_{ij} J_{ij} \mathbf{S}_i \cdot \mathbf{S}_j
\end{align}
Let us recall the algebraic properties of the spin operators:
\begin{align}
    [S^\alpha_i, S^\beta_j] &= i \delta_{ij}\epsilon^{\alpha\beta\gamma} \hbar S^\gamma_i,\\
    [S^\pm_i, S^z_j] &= \mp \delta_{ij} \hbar S^\pm_i,\\
    [S^+_i, S^-_j] &= 2\delta_{ij} \hbar S^z_i
\end{align}
The total spin operators are defined as:
\begin{align}
    \mathbf{S} &= \sum_i \mathbf{S}_i = \sum_i \big(S^x_i, S^y_i, S^z_i\big)^T,\\
    \mathbf{S}^2 &= \mathbf{S}\cdot \mathbf{S}
    = \Big(\sum_iS_i^x\Big)^2 + \Big(\sum_iS_i^y\Big)^2 + \Big(\sum_iS_i^z\Big)^2.
\end{align}
The raising and lowering operators are defined as:
\begin{align}
    S^\pm_i &= S^x_i \pm i S^y_i.
\end{align}
The total raising and lowering operators are defined as:
\begin{align}
    S^\pm &= \sum_i S^\pm_i.
\end{align}
The commutation relation between $H$ and $S^z$ is given by
\begin{align}
[H, S^z] &= \sum_{ij} J_{ij} [\mathbf{S}_i \cdot \mathbf{S}_j, S^z], \nonumber\\
&= \sum_{ij} J_{ij} \sum_k [\mathbf{S}_i \cdot \mathbf{S}_j, S^z_k].
\end{align}
Since
\begin{align}
    [S^x_iS^x_j, S^z_k] &= -\delta_{ik} S^y_i S^x_j - \delta_{jk} S^x_i S^y_j,\\
    [S^y_iS^y_j, S^z_k] &= \delta_{ik} S^x_i S^y_j + \delta_{jk} S^y_i S^x_j,\\
    [S^z_iS^z_j, S^z_k] &= 0,
\end{align}
then, $[H,S^z] = 0$. Generally,
\begin{align}
    [H,S^z] = [H,S^x] = [H,S^y] = 0.
\end{align}
Therefore, $H$ is SU(2) invariant. Moreover,
\begin{align}
[H,\mathbf{S}^2] = [H,{S^x}^2] + [H,{S^y}^2] + [H,{S^z}^2] =0\,.
\end{align}
\section{Fubini-Study Metric}
\label{app:block-diagonal-fubiny-study}
The infinitesimal Euclidean distance between two parameterized quantum states $\rho_{\boldsymbol{\theta}} = \ket{\psi(\boldsymbol{\theta})}\bra{\psi(\boldsymbol{\theta})}$ and $\rho_{\boldsymbol{\theta} + d\boldsymbol{\theta}} = \ket{\psi(\boldsymbol{\theta} + d\boldsymbol{\theta})}\bra{\psi(\boldsymbol{\theta} + d\boldsymbol{\theta})}$, such that $\boldsymbol{\theta}\in \mathbb{R}^n$ can be expressed as
\begin{align}
d^2(\rho_{\boldsymbol{\theta}}, \rho_{\boldsymbol{\theta} + d\boldsymbol{\theta}}) &= \sum_{(i,j) \in [n]^2} g_{ij}(\boldsymbol{\theta}) d\theta_i d\theta_j,
\end{align}
where $g_{ij}(\boldsymbol{\theta}) = \text{Re } G_{ij}(\boldsymbol{\theta})$, and the quantum geometric tensor $G_{ij}(\boldsymbol{\theta})$ is defined as~\cite{Stokes2020}:
\begin{align}
    G_{ij}(\boldsymbol{\theta}) &=
    \left\langle\frac{\partial \psi(\boldsymbol{\theta})}{\partial \theta_i}\frac{\partial \psi(\boldsymbol{\theta})}{\partial \theta_j}\right\rangle \nonumber\\
    &-
    \left\langle\frac{\partial \psi(\boldsymbol{\theta})}{\partial \theta_i}\psi(\boldsymbol{\theta})\right\rangle
    \left\langle\psi(\boldsymbol{\theta})\frac{\partial \psi(\boldsymbol{\theta})}{\partial \theta_j}\right\rangle.
\end{align}
This tensor is computed based on the Fubini-Study distance.\\
In the case of ansatz involving layers of parametrized Pauli rotation gates, the quantum geometric tensor can be
simplified analytically since the parameterized ansatz form is given by:
\begin{align}
 &U(\boldsymbol{\theta}) \ket{0} =\nonumber\\
 &V_L U_L(\boldsymbol{\theta}_L) V_{L-1}U_{L-1}(\boldsymbol{\theta}_{L-1}) \cdots V_1 U_1(\boldsymbol{\theta}_1) V_0 \ket{0},
\end{align}
such that $$U_l(\boldsymbol{\theta}_l) = \prod_{j=1}^N \exp(-i \theta_l^j P^j_l)$$ is the unitary operator of the $l$-th layer involving Pauli rotations on $N$ qubits, and $V_l$ is a fixed unitary operator that does not depend on the parameters $\boldsymbol{\theta}_l$.
Therefore, if the Fubini-Study metric is approximated by block matrices for each layer, the quantum geometric tensor can be approximated as $G_{ij}(\boldsymbol{\theta}) \approx \bigoplus_l G^l_{ij}(\boldsymbol{\theta}_l)$~\cite{Stokes2020}, where
\begin{align}
    \label{eq:block-diagonal-quantum-geometric-tensor}
    G^l_{ij}(\boldsymbol{\theta}_l) &=
\langle \psi_l \vert P^i_l P^j_l \vert \psi_l \rangle
- \langle \psi_l \vert P^i_l \vert \psi_l \rangle \langle \psi_l \vert P^j_l \vert \psi_l \rangle.
\end{align}
This quantity reflects the covariance of the two observables $P_l^i$ and $P_l^j$. According to this approximation, the blocks are symmetric matrices, and they are real valued. Moreover, each block requires at most $O(N^2)$ circuit executions since the second term in Eq.~\eqref{eq:block-diagonal-quantum-geometric-tensor} can be computed with single circuit in the $P_l^N \cdots P_l^1$ basis, while the first term can be computed at most with $O(N^2)$ circuits or with $O(1)$ circuits if the Pauli operations are the same for all gates.\\
However, the block matrices need to be invertible and significantly larger than zero to be meaningful. Actually, a small eigenvalued block matrix means a flat landscape. Therefore, it is good to design the ansatz in such a way that the blocks of the metric are not singular.\\
In case of the block corresponding to the first layer, generally, the qubits are initialized in tensor product state. Hence, the first block matrix is diagonal. To avoid singular first block, we can ensure that
$$ \langle \psi_1 \vert P_1^N \otimes \cdots \otimes P_1^1 \vert \psi_1 \rangle \neq 0.$$
This can be achieved by using suitable Pauli rotation generator according to the choice of the first layer $V_0$.\\
In our case, our ansatz was specifically designed to have a diagonal constant Fubini-Study metric. Namely,
\begin{align}
    \label{eq:full-quantum-geometric-tensor}
    G_{ij}(\boldsymbol{\theta}_l) &=
\langle \psi_l \vert P^i_l U^\dagger_{l,k-1} P^j_k \vert \psi_k \rangle
- \langle \psi_l \vert P^i_l \vert \psi_l \rangle \langle \psi_k \vert P^j_k \vert \psi_k \rangle,
\end{align}
where
\begin{align*}
    U_{l,k-1} =U^\dagger_l(\boldsymbol{\theta}_l)V^\dagger_l \cdots U^\dagger_{k-1}(\boldsymbol{\theta}_{k-1}) V^\dagger_{k-1},
\end{align*}
and ($l$, $k$) are the layer indices where ($i$, $j$) parameter indices are located.
We can exploit the following identities to analytically prove that the full metric is constant and diagonal:
\begin{align}
\vcenter{\hbox{\includegraphics[width=90pt]{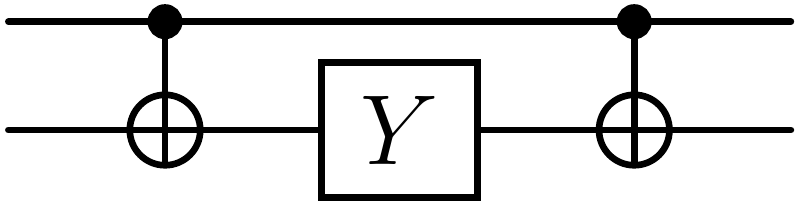}}}
&=
\vcenter{\hbox{\includegraphics[width=43pt]{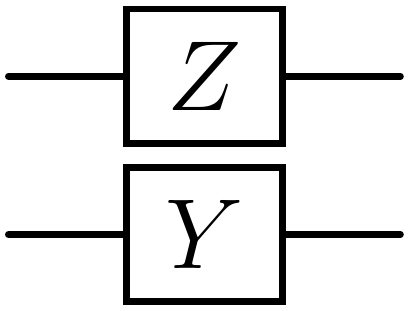}}}
\\
\vcenter{\hbox{\includegraphics[width=90pt]{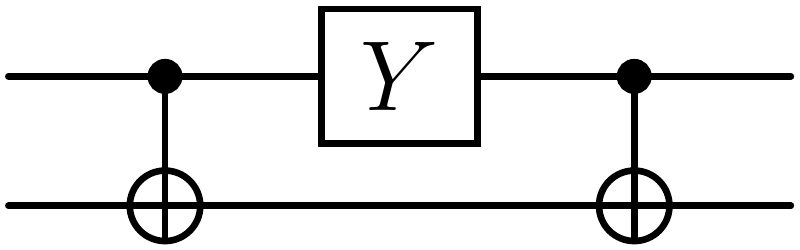}}}
&=
\vcenter{\hbox{\includegraphics[width=43pt]{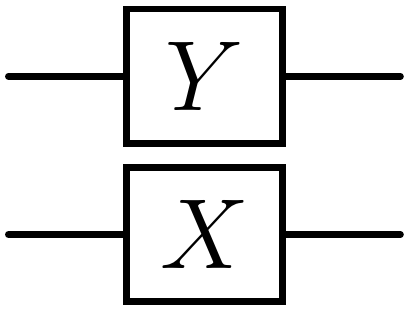}}}
\\
\vcenter{\hbox{\includegraphics[width=90pt]{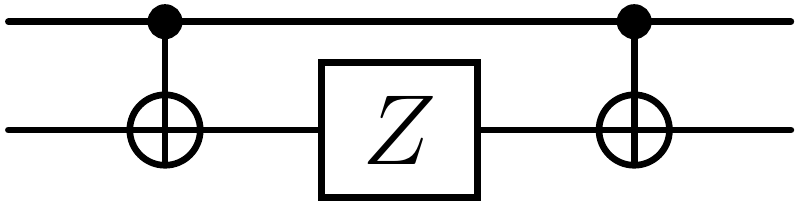}}}
&=
\vcenter{\hbox{\includegraphics[width=43pt]{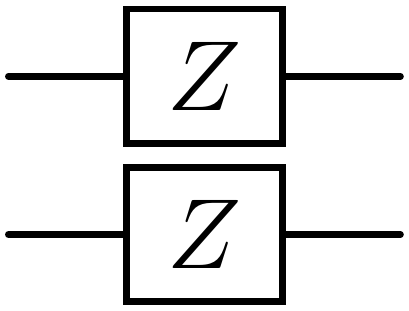}}}
\\
\vcenter{\hbox{\includegraphics[width=90pt]{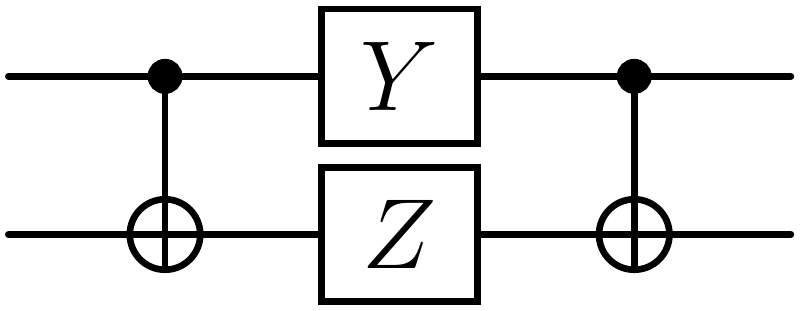}}}
&=
\vcenter{\hbox{\includegraphics[width=43pt]{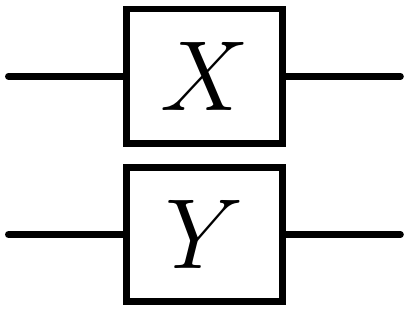}}}\\
\vcenter{\hbox{\includegraphics[width=90pt]{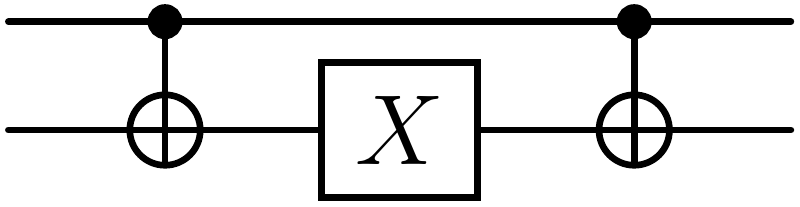}}}
&=
\vcenter{\hbox{\includegraphics[width=43pt]{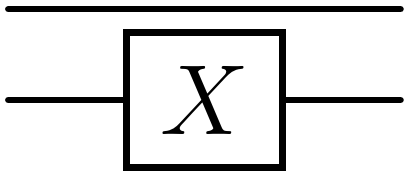}}}
\end{align}
In the end, the expectation values that involve single $Y_i$ gates or double $Y_iY_j$ gates in separate qubits can be simplified with these identities to remove the entangling gates leaving us with a term of the form $\lambda \times \bra{0} HYH\ket{0}$ which is equal to zero no matter what $\lambda$ is. Only the expectation values of $Y^2_i$ give unities.
\section{Qubit-Wise Commuting Pauli Terms Grouping Measurement}
\label{sec:qwc-pauli}
The Hamiltonian in Eq. \eqref{eq:hamiltonian} for one kagome star (12~qubits) yields 54 terms to be measured on a quantum computer. However, it is possible to optimize this number of evaluations by grouping qubit-wise commuting Pauli terms, since all commuting operators are simultaneously measurable. Namely,
\begin{align}
\left[ X_i X_j, X_k X_l\right]=\left[ Y_i Y_j, Y_k Y_l\right] =\left[ Z_i Z_j, Z_k Z_l\right] = 0
\end{align}
for any qubits $i,j,k,l$. Hence, it is possible to compute the 54 terms--or any number of Heisenberg interaction terms--by evaluating three quantum circuits only, namely,
$$\left\langle \psi \right| \sum_{\langle i, j \rangle} \sigma_i \sigma_{j}\left| \psi \right\rangle$$
for $\sigma=X, Y, Z$ such that $\psi$ is the trial wave function.
The effect of grouping commuting Pauli terms on the measurement accuracy has been studied in literature~\cite{Wecker2015,McClean2016,Romero2019,Crawford2021}. Interestingly, \citet{Crawford2021} showed that splitting a commuting group to two sets does not improve the accuracy under the condition that the number of measurements is optimally distributed among the groups. In this study, the standard error calculation, based on the variance of energy obtained from sampling the grouped commuting terms, is discussed in the following section.
\subsection{Measured Energy Variance}
\label{app:standard-error}
We show here how we compute the empirical variance and the standard deviation of energy value obtained from qubit-wise commuting Pauli words grouping. In the experiment, we showed that sampling three circuits in the $X$, $Y$, and $Z$ basis is enough to estimate the energy as follows:
\begin{align}
    \left\langle H \right\rangle &=
    \left\langle H_X \right\rangle +
    \left\langle H_Y \right\rangle +
    \left\langle H_Z \right\rangle,
\end{align}
such that\begin{align}
    \left\langle H_X \right\rangle &\approx \frac{1}{N} \sum_{b\in \mathcal{S}_X} \langle b\vert \sum_{\langle i,j\rangle} Z_iZ_j \vert b \rangle,\\
    \left\langle H_Y \right\rangle &\approx \frac{1}{N} \sum_{b\in \mathcal{S}_Y} \langle b\vert \sum_{\langle i,j\rangle} Z_iZ_j \vert b \rangle,\\
    \left\langle H_Z \right\rangle &\approx \frac{1}{N} \sum_{b\in \mathcal{S}_Z} \langle b\vert \sum_{\langle i,j\rangle} Z_iZ_j \vert b \rangle.
\end{align}
where $\mathcal{S}_X$, $\mathcal{S}_Y$, and $\mathcal{S}_Z$ are independent sets of bitstrings that are measured in the $X^{\otimes n}$, $Y^{\otimes n}$, and $Z^{\otimes n}$ eigenbases, respectively.
Statistically, since the sets of samples are independent, the covariance of the variables $H_X$, $H_Y$, and $H_Z$ is zero. Hence, the variance of $H$ is nothing but
\begin{align}
    \text{Var}(H) = \text{Var}(H_X) + \text{Var}(H_Y) + \text{Var}(H_Z).
\end{align}
The variance of each variable $H_P$ can be computed as follows
\begin{align}
    \text{Var}(H_P) =
    \frac{1}{N}\sum_{b\in \mathcal{S}_P} \langle b\vert \Big(\sum_{\langle i,j\rangle} Z_i Z_j - \langle H_P \rangle \Big)^2 \vert b \rangle,
\end{align}
which leads to the reduced form
\begin{align}
    \text{Var}(H_P) = \langle H_P^2 \rangle - \langle H_P \rangle^2.
\end{align}
We can show using the identity projection $I = \sum_{b\in \mathbb{Z}_2^n} \ket{b}\bra{b}$ that
\begin{align}
    \text{Var}(H_P) &=
    \frac{1}{N}\sum_{b\in \mathcal{S}_P} \Big(\langle b\vert \sum_{\langle i,j\rangle} Z_i Z_j \vert b \rangle - \langle H_P \rangle \Big)^2,
\end{align}
which is nothing but the empirical variance of the energy value obtained from the $P$ basis measurements. The standard deviation is then given by
\begin{align}
\label{eq:standard-deviation-app}
    \sigma(H) = \sqrt{\text{Var}(H)}.
\end{align}
The standard error is then
\begin{align}
\label{eq:standard-error-app}
    \epsilon = \sqrt{\frac{\text{Var}(H)}{N}}.
\end{align}
\section{Readout Error Mitigation}
\label{sec:rem}
One of the main sources of noise in quantum devices is the state preparation and measurement (SPAM) errors. Specifically, the measurement or readout errors are well-known in many areas of physics such as astronomy and high-energy physics. The main assumption is that such readout errors are governed by the response matrix $R$, which is a stochastic matrix that stores the probability of measuring a specific state $\ket{{f}} = \ket{f_0 f_1 \cdots f_N}$ given that the $N$ qubits are prepared in a prior state $\ket{{i}} = \ket{i_0 i_1 \cdots i_N}$. Here, $i, f \in \{0, 1\}^{\otimes N}$. Namely, the $2^N \times 2^N$ components of the $R$ matrix are
\begin{align}
    R_{{f}{i}} = P({f}|{i}),
\end{align}
which is interpreted as the probability of measuring $\ket{{f}}\bra{{f}}$ state given that the true state is $\ket{{i}}\bra{{i}}$. Therefore, the relation between the true and the noisy vectors of bitstrings $\mathbf{t} = \sum_k t_k \ket{k}\bra{k}$ and $\mathbf{n}= \sum_k n_k \ket{k}\bra{k}$ respectively are related as follows \begin{align}
\label{eq:response-matrix-application}
    n_k = \sum_l R_{kl} t_l,
\end{align}
which suggests that the true readout vector is nothing but $\mathbf{t} = R^{-1} \mathbf{n}$. Traditionally, Moore-Penrose pseudo-inverse is taken since the stochastic matrices are not generally invertible. Unfortunately, taking the inverse of an exponential matrix is a hard computation overhead. \citet{Benjamin2019} suggested unfolding techniques borrowed from high energy physics. However, the characterization of the full $R$ matrix by itself is not efficient since it requires $2^N$ state preparations, $\mathcal{O}(2^n)$ measurements on the QPU, and $2^n \times 2^n$ amount of memory for storage. Nevertheless, the full characterization of the $R$ matrix assumes correlated all-to-all interaction between the qubits.
\citet{Geller2021} proposed an efficient measurement method based on the characterization of the $k$ nearest neighbor qubits, requiring $O(4^k N^2)$ where $N$ is the number of qubits, but that leaves the burden of classically post-processing a $2^N\times2^N$ matrix which is still not scalable.
In practice, we can assume that correlated readout errors occur just locally between nearest neighbors in QPUs of restricted connectivity~\cite{Geller2021}, so
\begin{align}
&P(a_1a_2\cdots a_N|b_1b_2\cdots b_N)
&=\bigotimes_{\langle jk\cdots l\rangle\in \mathcal{P}} P_{jk\cdots l} (a|b),
\end{align}
such that $P_{jk\cdots l}$ is the probability of transition of the specific qubits labeled by $jk\cdots l$ sharing the same neighborhood, and $\mathcal{P}$ is the partition of the qubit register.
The immediate advantage of this model is the efficiency of taking the inverse of the response matrix since $P^{-1} = \bigotimes_{\langle jk\cdots l\rangle\in\mathcal{P}} P^{-1}_{\langle jk\cdots l\rangle}(a|b)$. Moreover, the correction process for a given measurement result $\boldsymbol{m} = \sum_k m_k \ket{k}\bra{k}$ involves only the application of the inverse partitioned response matrix on the relevant sections of the qubit register. That is, the recovered measurement result is
\begin{align}
    \boldsymbol{t} &= R^{-1} \sum_p m_p \ket{p}\bra{p}
    = \sum_p m_p R^{-1} \ket{p}\bra{p}.
\end{align}
We use the partitioned response matrix. Hence,
\begin{align}
    \boldsymbol{t} =
    \sum_p m_p \bigotimes_{\langle jk\cdots l\rangle}
    R^{-1}_{\langle jk\cdots l\rangle} \ket{p_{\langle jk\cdots l\rangle}}\bra{p_{\langle jk\cdots l\rangle}}.
\end{align}
We apply each $R^{-1}$ matrix partition to the corresponding $\ket{p}\bra{p}$ partition as a probability vector using the inverse of Eq.\eqref{eq:response-matrix-application}. That is,
\begin{align}
    \boldsymbol{t} =& \sum_p m_p \bigotimes_{\langle jk\cdots l\rangle}
    \sum_{q_{\langle jk\cdots l\rangle}}
    P^{-1}_{\langle jk\cdots l\rangle}(p|q) \ket{q_{\langle jk\cdots l\rangle}}\bra{q_{\langle jk\cdots l\rangle}}.
\end{align}
Since the number of measured samples of bitstrings in $\boldsymbol{m}$ should be small in practice, the global sum is still practical.
However, the tensor products can yield exponentially large samples. Let $N_p = N/P$ be the number of qubits, such that $P$ is the number of partitions. Then, the inner sum spans $2^{N_p}$ states, while the tensor product will yield $2^{N_p P}=2^{N}$ states in the computational basis. Nevertheless, it is possible to truncate results with the largest probabilities. It is worth noting that the REM method faces the issue of negative probabilities since the inverse does not preserve the stochasticity of the response matrix. Even though the negative pseudo-probabilities are absorbed in the estimation of any physical observable, it is important to note that the positivity of the pseudo-probabilities is crucial for maintaining the variational principle of Rayleigh and Ritz to be valid. Actually, positivity can be retained by involving more efforts to minimize $\vert\vert R \mathbf{t} - \mathbf{m}\vert\vert$~\cite{Geller2021} or minimize $\vert\vert \mathbf{t} - R^{-1} \mathbf{m}\vert\vert$. The latter is computationally less demanding. In this work, we show the REM results with both scenarios: the positivity preserving and non-preserving REM since the REM with positivity recovery is still valid given that, in practice, the negative pseudo-probability values have small magnitudes.
\section{Zero-Noise Extrapolation Error Mitigation}
\label{sec:zne}
One of the efficient error mitigation techniques is the zero-noise extrapolation, which was initially developed by \citet{Li2017} and \citet{Temme2017}. The basic idea behind this technique is the possibility to tune the quantum machine's noise with a single parameter $\lambda$ in such a way that the expectation value $\langle O \rangle$ depends on this error parameter as follows
\begin{align}
    \langle O \rangle(\lambda) = E_0 + \sum_{k=1}^n a_k \lambda^k + O(\lambda^{n+1}),
\end{align}
where $E_0$ is the expected value without noise. \citet{Li2017} showed that this is true when the evolution of the quantum circuit is modeled with the master equation $\frac{\partial \rho}{\partial t} = -[H, \rho] + \lambda \mathcal{L}[\rho]$ where $\mathcal{L}$ is the Lindblad superoperator. Although originally tuning $\lambda$ was performed by rescaling execution time in quantum computers~\cite{Li2017,Temme2017,Kandala2019,Kim2023}, it is practically preferred to apply a unitary circuit or gate folding on digital quantum devices~\cite{Dumitrescu2018,GiurgicaTiron2020,He2020}. Folding technique, despite its practicality, scales only the circuit internal decoherence noise, while in principle keeping the SPAM errors fixed.
In this work, unitary global folding is performed considering the folding points $1$, $3$ and $5$. Previously, different extrapolation methods have been used; \citet{Li2017} used the Richardson’s deferred approach, while \citet{GiurgicaTiron2020} benchmarked different methods on IBMQ London five-qubit chip, showing the instability of Richardson's extrapolation and better accuracy with exponential extrapolation.
The choice of the extrapolation method is still an ambiguous feature of ZNE method. Nevertheless, in this work, we choose the second-degree polynomial model since it fits the points minimally to avoid overfitting.
Moreover, to infer the standard error of the extrapolated zero-noise energy, we use the Bayesian polynomial regression (BPR) with heteroscedastic noise technique assuming a Gaussian process~\cite{Bishop2006, gelman2025}.
Lastly, it is worth noting that ZNE does not guarantee an upper bound of the exact energy and allows the violation of Rayleigh-Ritz variational principle since the method is not based on Hamiltonian measurement.
\label{sec:supplementary-materials}
\section{QPU Devices Technical Details}
\label{app:SI}
The native gates of Oslo IBM quantum device-which is featured by 7 qubits-are $\text{CNOT}$, $\text{RZ}$, $\sqrt{\text{X}}$, $\text{X}$, and the identity~$\text{I}$. Consequently, the required gates are implemented as follows:
\begin{align}
    \text{H} \equiv& \text{RZ}(\pi/2) \cdot \sqrt{\text{X}}\cdot \text{RZ}(\pi/2),\\
    \text{Ry}(\theta) \equiv&\text{RZ}(3\pi)\cdot \sqrt{\text{X}} \cdot\text{RZ}(\theta+\pi)\cdot\sqrt{\text{X}}.
\end{align}
The native gates of Kyoto IBM quantum device-which is from Eagle processor generation with 127 qubits-are $\text{ECR} = 1/\sqrt{2} (\text{IX}-\text{XY}) = \text{XI}\cdot \text{CNOT} \cdot \text{S}\sqrt{\text{X}}$, $\text{RZ}$, $\sqrt{\text{X}}$, $\text{X}$, and the identity~$\text{I}$. Therefore, the required gates are implemented as follows:
\begin{align}
    \text{H} \equiv& \text{RZ}(\pi/2) \cdot \sqrt{\text{X}}\cdot \text{RZ}(\pi/2),\\
    \text{Ry}(\theta) \equiv&\text{RZ}(3\pi)\cdot \sqrt{\text{X}} \cdot\text{RZ}(\theta+\pi)\cdot\sqrt{\text{X}},\\
    \text{CNOT} \equiv& \text{IX} \cdot \text{ECR} \cdot\\
    &\text{RZ}(-\pi)\cdot \sqrt{\text{X}} \cdot \text{RZ}(-\pi) \otimes \text{RZ}(-\pi/2).
\end{align}
The native gates of Torino IBM quantum device with the Heron architecture and 133 qubits are $\text{CZ}$, $\text{Rz}$, $\text{X}$, and~$\sqrt{\text{X}}$. The compiled gates in the Torino IBM quantum device with the Heron architecture are decomposed as follows:
\begin{align}
    \text{H} \equiv& \text{RZ}(\pi/2) \cdot \sqrt{\text{X}}\cdot \text{RZ}(\pi/2),\\
    \text{Ry}(\theta) \equiv&\text{RZ}(3\pi)\cdot \sqrt{\text{X}} \cdot\text{Rz}(\theta+\pi)\cdot\sqrt{\text{X}},\\
    \text{CNOT} \equiv& (I\otimes \text{RZ}({\pi}/{2}) \cdot{\sqrt{X}})\cdot \text{CZ} \cdot
    \nonumber\\ &\text{I} \otimes \text{RZ}(\pi) \sqrt{\text{X}} \cdot \text{RZ}({\pi}/{2}).
\end{align}
\bibliography{bibliography.bib}
\end{document}